\documentclass{article}

\usepackage{amsmath, amsfonts, mathtools, amsthm, geometry, subfiles, accents, todonotes, mdframed, graphicx, tcolorbox, subcaption, float}
\usepackage[utf8]{inputenc}
\usepackage{doi,url}

\usepackage{hyperref}
\usepackage{bookmark}
\usepackage{cleveref}

\newcommand{\pleb}{Pleba\'nski}
\newcommand{\sd}{\Sigma}

\newcommand{\R}{\mathbb{R}}

\newcommand{\cS}{\mathcal{S}}
\newcommand{\F}{\mathcal{F}}

\newcommand{\im}{{\rm i}}

\newcommand{\Tbox}{\Box\!\!\!\!{\scriptscriptstyle T}}

\newcommand\be{\begin{eqnarray}}
\newcommand\ee{\end{eqnarray}}

\newcommand{\tp}{\Phi}

\newtheorem*{theorem}{Theorem}

\title{K\"ahler decoupling for Kerr perturbations}
\author{
Stephen R. Green, Kirill Krasnov and Adam Shaw\\[1ex]
\textit{School of Mathematical Sciences, University of Nottingham, Nottingham, NG7 2RD, UK}
}

\begin{document}

\maketitle

\begin{abstract} The Euclidean Kerr metric is conformal, in two distinct ways, to a K\"ahler metric, with conformal factors determined by the repeated eigenvalue of the two chiral halves of the Weyl curvature. A Lorentzian analogue holds, where the conformally related metric is complex but retains key features of K\"ahler geometry. We show that this hidden K\"ahler structure provides a geometric explanation for the existence of decoupled equations for curvature scalars, such as the Teukolsky equations. The essential mechanism is that, on a K\"ahler background, self-dual 2-forms are parallel with respect to a natural covariant derivative, so differential operators acting on them preserve their decomposition and do not mix components. In this way, decoupling is seen to be a direct consequence of K\"ahler geometry. We make this mechanism explicit in two ways. First, we show that the spin-k Teukolsky operator can be obtained from a Laplace-type operator associated with the K\"ahler metric by a similarity transformation. Second, for electromagnetic perturbations, we use the conformal invariance of Maxwell’s equations $\delta \mathcal{F}=0$ to show that they imply $d \delta \mathcal{F}=0$, where $\delta$ is the co-differential of the K\"ahler metric. This operator automatically decouples, and the resulting equations for the extremal components coincide with the spin-one Teukolsky equations.
\end{abstract}

\tableofcontents

\section{Introduction}

In \cite{Teukolsky:1973ha} Teukolsky showed, using Newman-Penrose formalism, that suitably rescaled extreme spin weight Maxwell and linearised Weyl scalars on Kerr background satisfy decoupled and separable wave equations. Soon after Ryan \cite{Ryan:1974nt} showed that spin two Teukolsky equations can be obtained from the Penrose wave equation satisfied by the Riemann curvature in vacuum, i.e. an equation that holds when the Ricci part of the curvature vanishes
\begin{equation}
\square R_{\mu\nu\rho\sigma}
= R_{\mu\nu\alpha\beta} R_{\rho\sigma}{}^{\alpha\beta}
+ 2\left( R_{\mu\alpha \rho\beta} R_{\nu}{}^{\alpha}{}_{\sigma}{}^{\beta}
         - R_{\mu\alpha \sigma\beta} R_{\nu}{}^{\alpha}{}_{\rho}{}^{\beta} \right).
\end{equation}
To obtain the spin-two Teukolsky equation from the Penrose wave equation one linearises it and takes a suitable null tetrad projection. In \cite{Bini:2002jx} an analogous derivation was given for the spin-one Teukolsky equation, with the starting point being the Hodge-de Rham equation satisfied by the field strength 2-form. Moreover, it was observed in \cite{Bini:2002jx}, that the spin $k$ Teukolsky equation can be written very concisely as
\begin{equation}\label{spin-k-Teuk}
\mathcal{O}_k^T \Phi^{(k)}:= \left( \Tbox_{\,2k} - 4 k^2 \Psi_2 \right)\Phi^{(k)} = 0,
\end{equation}
where the introduced box operator is 
\begin{equation}\label{T-box}
\Tbox_{\,k} = g_{Kerr}^{\mu\nu} \left( \nabla^K_\mu + k T_\mu \right)
\left( \nabla^K_\nu + k T_\nu \right),
\end{equation}
and $T_\mu$ is a certain co-vector described explicitly in \cite{Bini:2002jx}, and $g_{Kerr}^{\mu\nu}, \nabla^K_\mu$ are the (inverse) Kerr metric and the Levi-Civita covariant derivative for Kerr respectively. 

These results give the Teukolsky equation a geometric interpretation. However, they do not explain the origin of the decoupling mechanism for the differently spin weighted scalars. If extremal weight scalars equations are simply projections of a covariant equation, can it be that the intermediate spin weight scalars also satisfy decoupled equations? And indeed, in the case of electromagnetic perturbations, it is known \cite{Fackerell:1972hg} that the spin weight zero scalar also satisfies a decoupled equation of its own, which is however non-separable. For spin two the situation is more complex. In \cite{Aksteiner:2010rh} it was argued that in the spin two case all intermediate spin weight scalars satisfy decoupled equations in suitable gauges. This was pursued further by \cite{Araneda:2016iwr}, who derived an equation for the spin weight zero scalar, see (1.25) of this paper. This equation, however, prior to a gauge is chosen, also depends on the metric perturbation. The geometric origin of the connection $T_\mu$ in (\ref{T-box}) also remained a mystery until \cite{Araneda:2018ezs}, where Araneda has shown that $T_\mu$, which he called the Teukolsky connection, is related to a distinguished conformally covariant connection. 

Araneda's results in \cite{Araneda:2016iwr}, \cite{Araneda:2018ezs} also make it clear that the essence of the decoupling is in introducing a suitable power of the background Weyl curvature scalar $\Psi_2$ in between two copies of the covariant derivative. This is very clear from the formulas (1.15)-(1.17) and (1.18)-(1.20) of \cite{Araneda:2016iwr}. A further observation was made by Green in the talk \cite{Green}, where the decoupling of spin-one perturbations is described in the differential forms formalism. The approach is developed into a Lorenz-gauge reconstruction scheme using two Debye potentials, including sources via a stream potential and gauge corrector. The main observation of \cite{Green} is that the equations
\begin{equation}\label{green}
\zeta^{-2} d \zeta^2 \delta \mathcal{F}=0,\qquad \zeta= \Psi_2^{-1/3},
\end{equation}
which are implied by the Maxwell's equations $\delta \mathcal{F}=0$, after projection onto the self-dual part of $\mathcal{F}$, form a decoupled system of equations for the field strength scalars. The key mechanism here is the introduction of the $\Psi_2^{-2/3}$ between the two derivatives. In the paper \cite{Araneda:2018ezs} it was understood that the essence of introduction of $\Psi_2^{-2/3}$ between the derivative has to do with the fact that Kerr metric is conformal to a Lorentzian K\"ahler metric. 

While these works identify important structures underlying the Teukolsky equations, the geometric origin of the decoupling mechanism remains somewhat implicit. In particular, the appearance of specific powers of the Weyl scalar $\Psi_2$ and the associated modified connection suggests an underlying structure, but it is not clear from this formulation why these modifications lead to decoupling, nor why the different spin components evolve independently. The main purpose of this paper is to show that this decoupling is a direct consequence of the Kähler geometry to which the Kerr metric is conformal. In our approach, the role of the conformal factor and the associated weights becomes transparent: the insertion of $\Psi_2^{-2/3}$ precisely converts the relevant operators into natural differential operators for the K\"ahler metric. The decoupling then follows immediately from a general property of K\"ahler geometry, namely that self-dual 2-forms are parallel with respect to a natural covariant derivative, so that differential operators acting on them preserve their decomposition.

We now summarise our main results. We first explain how the conformally related K\"ahler metric arises from the Kerr metric.  We then show that Teukolsky spin-k equations can be rewritten in terms of a natural Laplace-type operator for K\"ahler geometry, and then explain how K\"ahler geometry makes decoupling automatic. Finally, we demonstrate the role of K\"ahler geometry for the decoupling of Maxwell fields.

The existence of an integrable almost complex structure for type D spacetimes was discovered by Flaherty in \cite{Flaherty}. This was rediscovered in the context of four-dimensional Riemannian geometry by Derdzinski \cite{Derdzinski}. We recall that a 4-dimensional Riemannian manifold is said to be one-sided type D if two of the eigenvalues of a chiral half of its Weyl curvature coincide. We then have
\begin{theorem} {\bf (Derdzinski)} Let $(M,g')$ be a Riemannian 4-dimensional manifold that is Einstein and one-sided type D. Then $g=\Psi_2^{2/3} g'$ is a K\"ahler metric, where $\Psi_2$ is the repeated Weyl eigenvalue.
\end{theorem}
Derdzinski actually proves a stronger statement where one only needs to assume that the Weyl tensor of $g'$ is divergence-free, however the above is sufficient for our purposes. A simple proof of this theorem that is based on the Pleba\'nski formalism has been described by two of us in \cite{Krasnov:2024qyk}. The Euclidean Kerr metric is both-sided type D, which immediately implies that it is conformal to two different K\"ahler metrics. 

An analogous statement can be made in the context of Lorentzian geometry, with the caveat that in this case the integrable "almost-complex structure" operator $J$ is complex-valued. The two different (and commuting) integrable complex structures $J_\pm$ of the Euclidean Kerr become complex conjugates of each other for the Lorentzian Kerr. Similarly, the two different K\"ahler metrics in the Euclidean setting become complex and complex conjugates of each other in the Lorentzian setting. While the Lorentzian K\"ahler geometry is not as familiar as its Euclidean counterpart, the key features of Euclidean K\"ahler geometry continue to hold in the Lorentzian setting, and these are these features that are responsible for the decoupling. 

We now describe this complex Lorentzian K\"ahler metric more concretely. Let $r,q,t,\phi$ be the coordinates in which the Kerr metric takes the form (\ref{kerr-metric}). The coordinate $q$ is related to the usual $\theta$ via $q=\cos\theta$. The metric is algebraically special type D, which is manifested by the fact that the two chiral halves $W^\pm$ of the Weyl curvature tensor each have two of their eigenvalues coinciding. Concretely, each $W^\pm$ can be described as a $3\times 3$ matrix, which then takes the form of (\ref{Psi-Kerr}). Let us denote the cube root of (minus) the repeated eigenvalue 
\begin{align}
	\lambda_{\pm} = \left(\frac{\Psi_2^\pm}{M}\right)^{1/3}= \frac{1}{r\pm i a q}.
\end{align}
The conformally-related K\"ahler metric is given by
\begin{align}
g = \lambda_+^2 g_{Kerr}.
\end{align}
This metric is complex, and we could have introduced two such complex conjugate metrics via $g_\pm = \lambda_\pm^2 g_{Kerr}$. However, one of this metrics is sufficient for our story. The reason for this particular relation between the Einstein metric $g_{Kerr}$ and the K\"ahler metric $g$ becomes clear by following the proof of Derdzinski theorem, for a particularly simple argument explaining this particular conformal transformation see \cite{Krasnov:2024qyk}, formula (36). 

The fact that this K\"ahler geometry is behind Teukolsky equations is demonstrated by the following statement, which we prove in the main text, with some technical bits delegated to the Appendix. 
\begin{theorem} {\bf (A)} Define
\begin{equation}\label{O-k-intr}
\mathcal{O}_k = -(\nabla^\alpha + \im k a^\alpha) (\nabla_\alpha +\im k a_\alpha) + \frac{1+2k^2}{6} s,
\end{equation}
where $\nabla_\mu$ Levi-Civita connection of the K\"ahler metric, and $s$ is the K\"ahler metric scalar curvature. Let $\mathcal{O}^T_k$ be the Teukolsky spin $k$ operator from \cite{Teukolsky:1973ha}, see also (\ref{O-T}) below. Then 
\begin{equation}\label{comparison}
\mathcal{O}^T_k = \frac{\lambda_+}{\lambda_-} \frac{\lambda_+}{\Delta^{k/2}} \mathcal{O}_k \left( \frac{\Delta^{k/2}}{\lambda_+}\right),
\end{equation}
where $\Delta = r^2 - 2M r + a^2$.
 \end{theorem}
 This shows that Teukolsky spin-k equations can be rewritten in terms of a natural Laplace-type operator for K\"ahler geometry. This already hints at its relevance for the decoupling. In words,  the usual Teukolsky spin-k equations arise from the K\"ahler geometry equations by a similarity transformation combined with an additional final rescaling. Note that the additional rescaling factor can be written as $\lambda_+/\lambda_- = \rho^2 \lambda_+^2$, where $\rho^2=(\lambda_+\lambda_-)^{-1}=\sqrt{-{\rm det}(g_{Kerr})}$. So, the additional rescaling in the formula (\ref{comparison}) can be understood as the conformal transformation passing back to Kerr together with densitisation.
 
 The well-known properties of K\"ahler geometry now explain why the K\"ahler geometry Laplace-type equations do not mix scalars and exhibit decoupling. The most important for us property can be stated that there exists a complex-valued 1-form $a_\mu$ such that
\begin{equation}\label{derivative-omega}
\nabla_\mu \omega_{\rho\sigma} = 0, \qquad (\nabla_\mu \pm \im a_\mu) \Omega^\pm_{\rho\sigma} = 0.
\end{equation}
In words, the K\"ahler 2-form $\omega$ is parallel with respect to the Levi-Civita connection of the K\"ahler metric, and the 2-forms $\Omega^\pm$ are parallel with respect to $\nabla$ shifted by the connection $a_\mu$. In the Euclidean setting, the connection $a_\mu$ is the canonical Chern connection of the line bundles spanned by $\Omega^\pm$. Its ultimate origin is in the fact that (in the Euclidean setting) the self-dual ${\rm SU}_{SD}(2)$ connection reduces to a ${\rm U}(1)$ connection. An alternative way of phrasing this is to say that the holonomy of a Riemannian K\"ahler 4D manifold is special and lies in the group ${\rm U}(1)\times {\rm SU}_{ASD}(2)={\rm U}(2)$. The ${\rm SU}_{ASD}(2)$ here is the anti-self dual factor in the decomposition ${\rm SO}(4)={\rm SU}_{SD}(2)\times{\rm SU}_{ASD}(2)/\mathbb{Z}_2$. It is this reduction of ${\rm SU}_{SD}(2)$ to ${\rm U}(1)$ 
that ultimately underlies the decoupling. Thus, it is the fact that in K\"ahler geometry the self-dual 2-forms $\omega,\Omega^\pm$ are parallel with respect to the natural connection that underlies the decoupling of K\"ahler Laplace-type equations. Below we will explicitly demonstrate this on the example of spin-one, but the advocated here principle is general: {\it K\"ahler geometry implies decoupling}.

We now consider the case of spin-one perturbations of Kerr in detail. The fact that the K\"ahler metric is of importance for spin-one decoupling can be seen already from (\ref{green}). Indeed, we have
\[
d_K \zeta^2 \delta_K \sim d_K \lambda_+^{-2} \delta_K = - d \lambda_+^{-2} \star_K d \star_K,
\]
where $d_K$ is the exterior derivative, and $\delta_K$ is the Kerr metric co-differential and $\star_K$ is the Hodge star for the Kerr metric. The exterior derivative is metric independent $d_K=d$, and the Hodge star is conformally invariant on two-forms
\[
\star_K\Big|_{\Lambda^2} =  \star \Big|_{\Lambda^2},
\]
which is the rightmost occurrence of $\star_K$. The Hodge star on 3-forms is not conformally invariant and we have
\[
\lambda_+^{-2} \star_K\Big|_{\Lambda^3} =  \star \Big|_{\Lambda^3}.
\]
This means 
\[
d_K \lambda_+^{-2}  \delta_K = d \delta,
\]
where now $\delta$ is the co-differential for the K\"ahler metric. Thus, the operator that was observed in \cite{Green} to effect the decoupling of spin-one perturbations is just the K\"ahler geometry operator $d \delta$, which is a part of the K\"ahler Hodge Laplacian $\Delta_H = d\delta + \delta d$, with the last term being optional when acting on a closed 2-form $\mathcal{F}$. This already strongly suggests that the K\"ahler geometry is key to decoupling. 

We now explain in detail how K\"ahler geometry effects spin-one decoupling. Maxwell's equations in Kerr background read
\[
\delta_K \mathcal{F}=0.
\]
They are conformally invariant, and so they can instead be written as
\[
\delta \mathcal{F}=0,
\]
which in turn implies
\[
0= d\delta \mathcal{F} = (d\delta + \delta d)\mathcal{F} \equiv \Delta_H \mathcal{F},
\]
where $\Delta_H$ is the Hodge Laplacian of 2-forms. We then have the standard Weitzenbock identity for this Laplacian, see e.g. \cite{Brun-Weitz}, and also the Appendix
\[
(\Delta_H \mathcal{F})_{\mu\nu} = - \nabla^\alpha \nabla_\alpha \mathcal{F}_{\mu\nu} + \frac{s}{3} \mathcal{F}_{\mu\nu} - W_{\mu\nu}{}^{\rho\sigma} \mathcal{F}_{\rho\sigma},
\]
where the last term is the action of the Weyl curvature on the 2-form $\mathcal{F}_{\mu\nu}$, and $s$ is the scalar curvature. This operator can be restricted to the space of self-dual 2-forms, which for K\"ahler geometry is spanned by the following three 2-forms
\[
\Lambda^+ = {\rm Span}(\omega, \Omega^\pm).
\]
The form $\omega$ is the closed K\"ahler form, and $\Omega^\pm$ are $(2,0)$ and $(0,2)$ type forms. Expanding the self-dual part $\mathcal{F}_+$ of the field strength into the basis of self-dual 2-forms $\omega,\Omega^\pm$ we introduce the three scalars
\[
\mathcal{F}_+ = \Phi^+ \Omega^- + \Phi \omega + \Phi^+ \Omega^-.
\]
Using (\ref{derivative-omega}), as well as the specific form of the self-dual part of the Weyl curvature that holds in K\"ahler geometry, see (\ref{weyl-kahler}), one finds three decoupled differential equations for the scalars $\Phi,\Phi^\pm$
\[
\Delta_H \mathcal{F}_+ =0 \qquad \Leftrightarrow \qquad - \nabla^\alpha\nabla_\alpha \Phi =0, \qquad - (\nabla^\alpha \pm \im a^\alpha) (\nabla_\alpha \pm \im a_\alpha) \Phi^\pm + \frac{s}{2} \Phi^\pm =0.
\]
The most important point of the above discussion is that K\"ahler geometry guarantees decoupling because K\"ahler covariant derivative preserves the three subbundles of $\Lambda^+$ spanned by $\omega, \Omega^\pm$. The arising equations for $\Phi^\pm$ are then related to the Teukolsky spin-one equations by a similarity transformation, as is stated by Theorem A. 

While decoupling of K\"ahler geometry equations is automatic, there is no guarantee of separability. And indeed, the equation $\nabla^\alpha \nabla_\alpha \Phi=0$ is a version of the Fackerell-Ipser equation \cite{Fackerell:1972hg}, which is not separable. To exhibit separability of the equations for $\Phi^\pm$ a detailed study is necessary, which is carried out in the main text. Separability also follows from the fact, summarised by Theorem A, that the equations for $\Phi^\pm$ are related to the separable Teukolsky equations by a similarity transformation. 
  
 To prepare for our discussion of the spin-two case, we will summarise the spin-one results in terms of operators $d_+:\Lambda^1\to \Lambda^+$ and its adjoint rather than in terms of $d,\delta$. This operator arises naturally if one restricts the exterior derivative on 1-forms to self-dual 2-forms. We have
 \begin{theorem} {\bf (B)} Let $d_+:\Lambda^1 \to \Lambda^+$ be the restriction of the exterior derivative on 1-forms to the self-dual 2-forms, and let $d_+^*:\Lambda^+ \to \Lambda^1$ be its adjoint, computed with respect to the K\"ahler metric. Then $\mathcal{O} = 2 d_+ d_+^*= \Delta_H P_+$, where $P_+$ is the projector on $\Lambda^+$,  preserves each of the subspaces of $\Lambda^+$ spanned by $\omega, \Omega^\pm$, and its restriction $\mathcal{O}_\pm$ to $\Omega^\pm$ subspaces coincides with the operator $\mathcal{O}_{\pm 1}$ introduced in (\ref{O-k-intr}), and related to the spin-one Teukolsky operator via (\ref{comparison}). 
  \end{theorem}
 
 We have thus verified that K\"ahler geometry is at the root of decoupling for spin-one perturbations. Theorem A together with the general advocated principle (K\"ahler $\Rightarrow$ decoupling) suggests that this must also work for spin-two perturbations. Without a detailed study of the much more involved spin-two case, which is beyond the scope of this paper, we can only attempt a guess at the relevant operator. In view of Theorem B, and motivated by the Pleba\'nski formalism that we will be using in the main text, a natural spin-two guess would be to consider 
 \[
 d_+^{(2)} :\Lambda^1 \otimes \Lambda^+ \to S^2_0(\Lambda^+), 
 \]
 together with its adjoint $(d_+^{(2)})^*$. Then $ d_+^{(2)} (d_+^{(2)})^*$ is a natural Laplace-type operator on the space $S^2_0(\Lambda^+)$ of symmetric tracefree endomorphisms of $\Lambda^+$, where the perturbation of the self-dual part of the Weyl curvature lives. According to our general principle that K\"ahler implies decoupling, this operator will preserve the 5 different one-dimensional subspaces of $S^2_0(\Lambda^+)$. However, an explicit computation shows that the restriction of this operator to the extreme weight scalars is {\it not} the spin-two Teukolsky. Instead one gets
 \begin{equation}\label{spin-2-intr}
 d_+^{(2)} (d_+^{(2)})^* \Big|_{\Omega^\pm\otimes\Omega^\pm} = -(\nabla^\alpha \pm 2\im a^\alpha) (\nabla_\alpha \pm 2\im  a_\alpha) + s.
 \end{equation}
 Thus, while the derivative part of the arising this way operator is correct, the coefficient in front of the scalar curvature in this operator is the wrong (1 instead of 3/2) to match the Teukolsky spin-two operator. The arising operator is not separable because its potential term does not separate. However, it is possible that some modification of this construction works. We make some more comments on the spin-two case in the discussion section, but delegate complete treatment to a separate publication. 

The organisation of this paper is as follows. For the benefit of the reader, we review aspects of Riemannian K\"ahler geometry in the next section. Section \ref{sec:kerr} describes the Kerr metric, together with its \pleb{} description that will be useful for some computations. Here we also introduce the conformally related K\"ahler metric, and collect some number of useful results about it, with some more useful identities proven in the Appendix. In section \ref{sec:teuk} we compare the spin-k Teukolsky Master Equation operator with the Laplace-type operator for K\"ahler geometry, and show by an explicit computation that they are related by a similarity transformation. Section \ref{sec:spin-one} treats the spin-one case. Here we compute the restriction of the Hodge Laplacian one 2-forms to the subspaces spanned by $\Omega^\pm$ and show that this gives the K\"ahler Laplace-type operator introduced in the previous section. We also describe the spin-two computation that leads to (\ref{spin-2-intr}). We conclude with a discussion.

\section{Aspects of K\"ahler geometry in 4D}
\label{sec:kahler}

Euclidean version of the Kerr metric is known to be conformal to two different K\"ahler metrics. Thus, K\"ahler geometry becomes relevant. Here we review some aspects of Riemannian signature K\"ahler geometry. For more details on this important subfield of differential geometry see e.g. \cite{Moroianu}. We will later develop analogous viewpoint on Lorentzian metrics. 

A K\"ahler geometry $(M,g,\omega)$ consists of a manifold $M$, a Riemannian metric $g$ and a closed 2-form $\omega$, such that $J_\mu{}^\nu$ defined via $J_\mu{}^\nu = \omega_{\mu\rho} g^{\rho\nu}$ is an integrable orthogonal complex structure 
\begin{align}
	 J^2 = -\mathbb{I} ,\quad {\rm and} \quad g_{\mu\nu} J_\rho{}^\mu J_\sigma{}^\nu = g_{\rho\sigma}, \quad {\rm and} \quad \nabla J = 0,
\end{align}
where $\nabla$ is the Levi-Civita connection for the metric $g$. 

In four dimensions we can encode the metric into a triple of self-dual 2-forms, as in \pleb{}  formalism to be reviewed below. One of these 2-forms can be chosen to be the real 2-form $\omega$, while the other two can be chosen to be complex conjugates of each other, to which we refer as $\Omega^+,\Omega^-$. The integrability condition implies \begin{align}\label{nabla-omega}
	\nabla_\mu \omega_{\rho\sigma} = 0.
\end{align}

In four dimensions the Levi-Civita $\mathfrak{so}(4)=\mathfrak{su}(2)\oplus \mathfrak{su}(2)$ value connection splits into its self- and anti-self dual parts. As is standard in K\"ahler geometry, we will work in the orientation in which $\omega$ is self-dual. Then the self-dual part of the Levi-Civita connection on a four-dimensional K\"ahler manifold is extremely simple, and is essentially a $\mathfrak{u}(1)$ connection, called the Chern connection. It is best described by the following equations
\begin{align}\label{d-omega-Omega}
	d \omega = 0, \quad d\Omega^\pm = \pm \im  a \wedge \Omega^\pm.
\end{align}
The sign in the second relation is different from the one in (\ref{derivative-omega}), but in this section we use the standard K\"ahler conventions, and in the rest of the paper we use the sign as in (\ref{derivative-omega}). In fact, more is true and we have
\begin{align}\label{nabla-Omega}
	 \nabla_\mu \Omega^\pm_{\rho\sigma} = \pm \im  a_\mu \Omega^+_{\rho\sigma}.
\end{align}

The endomorphism $J$ gives rise to a splitting on the space of 1-forms $ \Lambda^1 = \Lambda^{1,0} \oplus \Lambda^{1,0}$, where $\Lambda^{1,0}$ is in the $+\im$ eigenvalue subspace and $\Lambda^{0,1}$ is in the $-\im$ eigenspace. Using the integrability we know there are complex coordinates $z_1,z_2,\bar{z}_1,\bar{z}_2$ such that
\begin{align}
	dz_1, dz_2 \in \Lambda^{1,0}, \quad d\bar{z}_1, d\bar{z}_2 \in \Lambda^{0,1}.
\end{align}
Integrability also implies that $d = \partial + \bar{\partial}$, where $\partial : \Lambda^{p,q} \rightarrow \Lambda^{p+1,q}$ and $\bar{\partial} : \Lambda^{p,q} \rightarrow \Lambda^{p,q+1}$. The complex structure also induces a splitting on the space of 2-forms $\Lambda^2 = \Lambda^{2,0} \oplus \Lambda^{1,1} \oplus \Lambda^{0,2}$, the first factor in the superscript counts the number of $dz_1$ or $dz_2$ factors and the second counts the $d\bar{z}_1, d\bar{z}_2$ factors. Further, the space $\Lambda^{1,1}$ splits into its part spanned by $\omega$ and the irreducible part that we shall refer to as $\Lambda^{1,1}_0$. We then have the following facts
\begin{align}
\Lambda^- = \Lambda^{1,1}_0, \qquad \Lambda^+ = \mathbb{R} \omega \oplus [[ \Lambda^{2,0} ]], \qquad \Lambda^{2,0} = {\rm Span}_\mathbb{C}(\Omega^+).
\end{align}
Here $[[ \Lambda^{2,0} ]]$ denotes the real vector space underlying the 1-dimensional complex space spanned by $\Omega^+$. Concretely, elements of $[[ \Lambda^{2,0} ]]$ are $\mu \Omega^+ + \bar{\mu} \Omega^-, \mu\in \mathbb{C}$. 

If we write the K\"ahler form as
\be
\omega = \frac{1}{\im}\left(g_{1\bar{1}} dz^1 \wedge d\bar{z}^1 + g_{1\bar{2}} dz^1\wedge d\bar{z}^2 + g_{2\bar{1}} dz^2\wedge d\bar{z}^1 + g_{2\bar{2}} dz^2 \wedge d\bar{z}^2\right)
\ee
then 
\be
\frac{1}{2} \omega \wedge \omega = {\rm det}(g_{i\bar{j}}) dz^1  \wedge dz^2 \wedge d\bar{z}^1 \wedge d\bar{z}^2. 
\ee
Because in four dimensions the spaces $\Lambda^{2,0}$ and $\Lambda^{0,2}$ are complex 1-dimensional, we have 
\begin{align}
	\Omega^+ \sim dz_1 \wedge dz_2, \quad \Omega^- \sim d\bar{z}_1 \wedge d\bar{z}_2.
\end{align}
We can choose 
\begin{align}
	\Omega^+ = ({\rm det}(g_{i\bar{j}}))^{1/2} dz_1 \wedge dz_2, \quad \Omega^- =({\rm det}(g_{i\bar{j}}))^{1/2}  d\bar{z}_1 \wedge d\bar{z}_2,
\end{align}
so that $\Omega^+ \wedge \Omega^- \sim \omega^2$. Then
\begin{align}
	d \Omega^+ =  \frac{1}{2} d\log( {\rm det}(g_{i\bar{j}})) \wedge \Omega^+ = \frac{1}{2} \bar{\partial} \log( {\rm det}(g_{i\bar{j}})) \wedge \Omega^+.
\end{align}
To write the last equality we used $d=\partial+\bar{\partial}$ and the fact that $\Omega^+\in \Lambda^{2,0}$. 
Comparing with (\ref{d-omega-Omega}) we get
\begin{align}
	a = \frac{\im}{2} \left(\partial \log( {\rm det}(g_{i\bar{j}}))-\bar{\partial} \log( {\rm det}(g_{i\bar{j}})) \right).
\end{align}
This can also be written as
\begin{align}\label{a-kahler}
	a_\mu =  \frac{1}{2} J_\mu{}^\nu \partial_\nu \log( {\rm det}(g_{i\bar{j}})) \equiv \frac{1}{2}\omega_\mu{}^\nu \partial_\nu \log( {\rm det}(g_{i\bar{j}})), \quad a = \frac{1}{2} \omega \cdot d\log( {\rm det}(g_{i\bar{j}})),
\end{align}
where we have introduced the operator ${(\omega \cdot \theta)}_\mu =\omega_\mu{}^\nu \theta_\nu$. As an immediate consequence of this we have 
\begin{align}
	\delta a = 0, 
\end{align}
where $\delta$ is the codifferential that appears as the adjoint of $d$ with respect to the inner product defined with the K\"ahler metric ${(\cdot,\cdot)} : \Lambda^p \otimes \Lambda^p \rightarrow \R$. 

The exterior derivative 
\be
\rho=da= -\im \partial \bar{\partial} \log( {\rm det}(g_{i\bar{j}}))
\ee
 is called the K\"ahler form. It has the decomposition
\begin{align}\label{s-4}
\rho = \frac{s}{4} \omega + \rho_0, \qquad \rho_0 \in \Lambda^- = \Lambda^{1,1}_0.
\end{align}
This means that for a K\"ahler metric it is easy to extract the scalar curvature $s$ from $da$. 

Further, for a K\"ahler metric in four dimensions, the self-dual part of the Weyl curvature tensor takes a very specific form. This directly follows from the fact that the self-dual part of the Levi-Civita connection, whose curvature encodes the self-dual part of the Weyl curvature together with the Ricci tensor, takes a very simple form. As we already described, while for a generic Riemannian 4D metric this self-dual part of the Levi-Civita is a $\mathfrak{su}(2)$ connection, for K\"ahler metric this is just a $\mathfrak{u}(1)$ connection that we denoted by $a$ above. This simplicity of the connection results in the fact that the Ricci tensor is completely encoded in the exterior derivative of $a$, and a very specific form of the self-dual part of Weyl. To describe the self-dual Weyl curvature $W_+$ we select a basis in $\Lambda^+$ appropriately and describe $W_+$ in matrix form. As before, we choose $\Omega^1=\omega$, and $\Omega^\pm = (\Omega^2\pm \im \Omega^3)/\sqrt{2}$, so that $\Omega^\pm$ span the spaces $\Lambda^{2,0}, \Lambda^{0,2}$ respectively. With respect to the basis $\Omega^{1,2,3}$ the self-dual part of the Weyl curvature takes the form of the following diagonal $3\times 3$ matrix
\begin{align}\label{weyl-kahler}
W_+ = \left( \begin{array}{ccc} \frac{s}{6} & 0 & 0 \\ 0 & -\frac{s}{12} & 0 \\ 0 & 0 & -\frac{s}{12} \end{array}\right).
\end{align}
This form of $W_+$ follows from the fact that the sum $(s/12)\mathbb{I} + W_+$ should reproduce the first term on the right-hand side of (\ref{s-4}). In matrix notations
\be
\frac{s}{12} \mathbb{I} + W_+ =  \left( \begin{array}{ccc} \frac{s}{4} & 0 & 0 \\ 0 &0 & 0 \\ 0 & 0 & 0 \end{array}\right).
\ee
All these properties, being algebraic, translate to Lorentzian K\"ahler metrics and will be useful for the computations to follow.

\section{Kerr geometry and its \pleb{} description}
\label{sec:kerr}

\subsection{\pleb{} Formulation}

We will be using the \pleb{} formulation of general relativity for some of the calculations, and we will start by reviewing this. We are very brief here, for more details the reader can consult e.g. \cite{Krasnov:2010olp} or the book \cite{Krasnov:2020lku}. The basic objects of the \pleb{} formalism are $\Sigma^i, A^i$ and $\Psi^{ij}$ which are a 2-form, 1-form and 0-form respectively. The field equations of the Lorentzian version of this formalism are
\begin{align}
	d \Sigma^i + \epsilon^{i j k} A^j \wedge \Sigma^k & = 0, \label{compat} \\
	F^i = \left(\frac{s}{12} \delta^{ij}+ \Psi^{ij}\right) \Sigma^j, & \label{eq:curvature-cond} \\
	\Sigma^i \wedge \Sigma^j = \frac{1}{3} \Sigma^k \wedge \Sigma^k \delta^{i j}, & \\
	\Sigma^i \wedge \bar{\Sigma}^j = 0, \quad {\rm Re}(\Sigma^i \wedge \Sigma^i) & = 0. \label{eq:reality-cond} 
	\end{align}
In the above equations $F^i = dA^i + \frac{1}{2} \epsilon^{ijk} A^j \wedge A^k$ is the curvature of the connection $A^i$, $s$ is the scalar curvature, and $\Psi^{ij}$ is the $3\times 3$ matrix encoding the self-dual half of the Weyl curvature. When the third equation is satisfied the 2-forms $\Sigma^i$ define a metric in which they are self-dual. The bottom two conditions then impose that the metric is real-valued, the resulting metric can be recovered using the Urbantke formula,
\begin{align}
	\sqrt{-g} g_{\mu\nu} = \frac{\im}{12} \epsilon^{ijk} \Sigma^i_{\mu\rho} \Sigma^j_{\nu\sigma} \Sigma^k_{\alpha\beta} \tilde{\epsilon}^{\rho\sigma\alpha\beta}. \label{eq:urbantke}
\end{align}
Here, $\tilde{\epsilon}^{\mu\nu\rho\sigma} = \pm 1$ is the totally antisymmetric symbol. It can be shown that the 2-forms $\Sigma^i_{\mu\nu}$ with one of their indices raised using the metric (which $\Sigma$'s define) satisfy the algebra of the imaginary quaterions
\begin{align}	\label{quat-algebra}
	\Sigma^i_\mu{}^\rho \Sigma^j_\rho{}^\nu = -\delta^{ij} \delta_\mu^\nu + \epsilon^{ijk} \Sigma^k_\mu{}^\nu.
		\end{align}
The first equations (\ref{compat}) implies that $A^i$ is the self-dual half of the Levi-Civita connection. Finally, (\ref{eq:curvature-cond}) becomes the Einstein's equation on the metric.

\subsection{Kerr metric}

We start by writing the Kerr metric in a form that makes the choice of the orthonormal coframe obvious. It is convenient to treat $q=\cos(\theta)$ rather than $\theta$ as the coordinate. The metric takes the following form
\begin{align}\label{kerr-metric}
 g_{Kerr} = -\frac{C}{r^2+a^2 q^2}{(dt - a(1-q^2)d\phi)}^2 + \frac{r^2+a^2 q^2}{C} dr^2 +\frac{r^2+a^2 q^2}{D} dq^2 + \frac{D}{r^2+q^2}{(a dt - (r^2+a^2)d\phi)}^2,
\end{align}
where $C = C(r), D = D(q)$ are polynomials in their respective variables
\begin{align}\label{C-D}
	C(r) = r^2 - 2Mr + a^2, \quad D(q) = 1-q^2  = \sin(\theta)^2,
\end{align}
where $M$ is the mass of the black hole and $a$ is the angular parameter. While this choice of $C(r), D(q)$ gives Kerr, there is a more general \pleb-Demianski class \cite{Plebanski:1976gy} of solutions of GR that can be obtained for other choices of $C,D$. It is convenient to keep $C,D$ general, specialising to (\ref{C-D}) when necessary.
The choice of the coframe is obvious
\begin{alignat}{2}\label{frame}
	e^0 &= \frac{\sqrt{C}}{\sqrt{r^2+a^2 q^2}} (dt - a(1-q^2)d\phi) ,\quad &&e^1 = \frac{\sqrt{r^2+a^2 q^2}}{\sqrt{C}} dr, \\
	e^2 &= \frac{\sqrt{r^2+a^2 q^2}}{\sqrt{D}} dq,\quad &&e^3 = \frac{\sqrt{D}}{\sqrt{r^2+a^2 q^2}} (a dt - (r^2 + a^2) d\phi).
\end{alignat}
For later reference we note that the associated volume form is
\begin{align}\label{volume-form}
e^0 \wedge e^1 \wedge e^2 \wedge e^3 = - (r^2+ a^2 q^2) dt \wedge dr \wedge dq\wedge d\phi.
\end{align}

\subsection{Chiral Objects}
Given the frame (\ref{frame}) we can introduce the self-dual 2-forms
\begin{align}
	\Sigma^1 &= \im (dt - a(1-q^2)d\phi) \wedge dr - dq \wedge ((r^2+a^2)d\phi - a dt), \\
	\Sigma^2 &= \frac{\im \sqrt{C}}{\sqrt{D}} ( dt - a(1-q^2) d\phi ) \wedge dq - \frac{\sqrt{D}}{\sqrt{C}} dr \wedge (a dt + (r^2 + a^2) d\phi)), \\
	\Sigma^3 &= \im \sqrt{C} \sqrt{D} dt \wedge d\phi - \frac{r^2 + a^2 q^2}{\sqrt{C} \sqrt{D}} dr \wedge dq.
\end{align}
By solving the equations $d \Sigma^i + \epsilon^{ijk} A^j \wedge \Sigma^k = 0$, we find the connections to be, see \cite{Krasnov:2024qyk} for details
\begin{align}
	A^1 &= \frac{1}{r^2+a^2 q^2} \left( \im C \partial_r \log\left( \frac{\sqrt{C}\sqrt{D}}{r+\im a q} \right)(dt - a(1-q^2)d\phi) + D \partial_q \log\left(\frac{\sqrt{C}\sqrt{D}}{r+\im aq} \right) \right)((r^2+a^2)d\phi - a dt), \\
	A^2 &= -\frac{\sqrt{C}\sqrt{D}}{r-\im a q} d\phi, \\
	A^3 &= \frac{-\im \sqrt{C} \sqrt{D}}{r+\im a q} \left( \frac{a}{C} dr + \frac{\im}{D} dq \right).
\end{align}
Another, more useful, way of writing these connections is as follows. 
Using the null internal vectors (\ref{L-S}) we can write the connections as
\begin{align}\label{Kerr-connection}
	A^i = \Sigma^i \cdot d \log \left( r+\im a q\right) + L^i a, \quad a =  \Sigma^1 \cdot d \log \left( \frac{\sqrt{CD}}{(r + \im a q){}^2}\right).
\end{align}

Computing the curvature of these connections we confirm that Einstein's equations are satisfied
\begin{align}
	F^i = \Psi^{ij} \Sigma^j
\end{align}
where
\begin{align}\label{Psi-Kerr}
	\Psi^{ij} = \begin{pmatrix}
		 2 \Psi_2 & 0 & 0 \\ 0 & - \Psi_2 & 0 \\ 0 & 0 & -\Psi_2
	\end{pmatrix}, \quad  \Psi_2= \frac{M}{(r+\im a q){}^3}.
\end{align}
We can write the resulting matrix $\Psi^{ij}$ using the internal null vectors $L^i,S^i,\bar{S}^i$ introduced in the Appendix
\begin{align}
	\Psi^{ij} = 2 \Psi_2 M^{ij} = 2 \Psi_2 (L^i L^j - S^{(i} \bar{S}^{j)}).
\end{align}

\subsection{Kerr metric is conformal to a (complex Lorentzian) K\"ahler metric}

We now start preparing for the main calculation, which will be carried out in the background of a metric that is conformal to Kerr metric. It is know, see e.g. \cite{Krasnov:2024qyk} for a recent exposition, that Euclidean Kerr metric is conformal (in two different ways) to K\"ahler metrics. The analogous statement is true for the Lorentzian Kerr, but in this case the conformal factor is complex, and so the arising "K\"ahler" metric is complex and Lorentzian. Nevertheless, most of the simplifying features of K\"ahler geometry continue to hold for this metric. In particular, the covariant derivative (with respect to the Levi-Civita connection) of the self-dual 2-forms take a very simple form characteristic of K\"ahler geometry. To put it differently, in K\"ahler geometry the part of the Levi-Civita connection that is needed to act on self-dual 2-forms reduces to a ${\rm U}(1)$ connection, and this continues to be true in the considered here case. 

Motivated by this discussion, we introduce
\begin{align}
	\lambda_{\pm} = \frac{1}{r\pm i a q},
\end{align}
which are clearly related to the cubic root of the $\Psi_2$ and $\overline{\Psi_2}$ components of the Weyl curvature. We then define
\begin{align}
g = \lambda_+^2 g_{Kerr}, \quad g^{-1} = \lambda_+^{-2} g_{Kerr}^{-1}.
\end{align}
The introduced metric $g$ is complex-valued, but it is conformal to a real Lorentzian Kerr metric. As we have already discussed, the reason for introducing this metric is that many of the algebraic statements that are true in K\"ahler geometry continue to be true for the complex metric $g$. Note that, unless explicitly stated to the contrary, all the calculations in this paper are done using the metric $g$ rather than $g_{Kerr}$.

Note that we have only defined one (complex) K\"ahler  metric $g$. But with $\lambda_-=\overline{\lambda_+}$ it is clear that another relevant K\"ahler  metric is just the complex conjugate of $g$. In Riemannian signature the two conformal factors $\lambda_\pm$ are real, and one obtains two genuine Riemannian signature K\"ahler metrics by this procedure. In Lorentzian signature one instead obtains a single complex-valued "Lorentzian" K\"ahler metric. We will not attempt the formalise the meaning of the "Lorentzian" K\"ahler here, some additional discussion on this is available in the papers \cite{Araneda:2018ezs}, \cite{Aksteiner:2022bwr}. It is sufficient for our purposes to understand that this is a metric that is conformal to a real Lorentzian (Kerr) metric, and also shares some key identities that are true in Riemannian K\"ahler geometry. 

We will now introduce the \pleb{} 2-forms for the metric $g$. These are related to those for the Kerr metric via
\begin{gather}
	\Omega^i_{\mu\nu} = \lambda_+^2 \Sigma^i_{\mu\nu}, \quad \Omega^i_\mu{}^\nu = \Sigma^i_\mu{}^\nu, \quad \Omega^{i\mu\nu} = \lambda_+^{-2} \Sigma^{i\mu\nu}.
\end{gather}
The indices of the objects related to the K\"ahler metric $g$ are raised and lowered with $g$, whereas the Kerr object indices are raised and lowered with $g_{Kerr}$. It is clear that the introduced objects have the property the endomorphisms $\Omega^i_\mu{}^\nu$ continue to satisfy the algebra of imaginary quaterions
\begin{align}\label{omega-algebra}
	\Omega^i_\mu{}^\rho \Omega^j_\rho{}^\nu = -\delta^{ij} \delta_\mu^\nu + \epsilon^{ijk} \Omega^k_\mu{}^\nu. \quad 
	\end{align}
Another key simplification is that after the conformal transformation the connection $a^i$ defined by $\Omega^i$ via 
\begin{align}
d\Omega^i + \epsilon^{ijk} a^j\wedge \Omega^k=0
\end{align}
is extremely simple. Indeed, using the result that a conformal transformation $g\to \lambda^2 g$ affects the connections via $A^i \rightarrow A^i + \Sigma^i \cdot d\log(\lambda)$, as well as the expression (\ref{Kerr-connection}) for the Kerr connection, it is clear that the chiral connection defined by $\Omega^i$ is 
\begin{align}
a^i = L^i a,
\end{align}
where $a$ is given by (\ref{Kerr-connection}), and $L^i$ is one of the vectors introduced in the Appendix. This is the analog of the usual K\"ahler geometry formula (\ref{a-kahler}). 

More conveniently, we introduce 
\be
\Omega^1=\omega\qquad \Omega^\pm=\frac{1}{\sqrt{2}} (\Omega^2\pm \im \Omega^3).
\ee 
We then have 
\begin{align}\label{nabla-omega-Omega}
\nabla_\mu \omega_{\rho\sigma}=0, \qquad (\nabla_\mu \pm \im a_\mu) \Omega^\pm_{\rho\sigma}=0. 
\end{align}
This simplicity of the covariant derivatives of the self-dual 2-forms $\omega,\Omega^\pm$ is the main reason to pass and do all the calculations in terms of these objects.

The fact that only one of the components of the chiral connection $a^i$ is non-zero directly implies that the self-dual part of the Weyl curvature has a very specific form (\ref{weyl-kahler}). Indeed, in general we have
\be
F^i(a) = \frac{s}{12} \Omega^i + W_+^{ij} \Omega^j + \text{terms in } \Lambda^-,
\ee
where $s$ is the scalar curvature. But $F^1= da$ and $F^2=F^3=0$, which directly implies that the matrix $W_+^{ij}$ is of the form (\ref{weyl-kahler}). This also implies that
\be\label{da}
da = \frac{s}{4} \omega + \text{terms in } \Lambda^-.
\ee
Contracting this with $\omega$ and using $(\omega,\omega) = (1/2) \omega^{\mu\nu} \omega_{\mu\nu}=2$ gives
\be
\frac{s}{2}= (da,\omega) = \frac{1}{2} \omega^{\mu\nu} ( \partial_\mu a_\nu - \partial_\nu a_\mu) = \omega^{\mu\nu} \partial_\mu a_\nu.
\ee
Using the second formula in (\ref{Kerr-connection}), which we will write as
\be\label{a-f}
a_\mu = \omega_\mu{}^\nu \partial_\nu \log(\lambda_+ f), 
\ee
where we introduced
\be\label{f}
\qquad f := \sqrt{CD}\lambda_+,
\ee
we get
\be\label{s-delta-f}
s= 2\omega^{\mu\nu} \partial_\mu (  \omega_\nu{}^\rho \partial_\rho \log(\lambda_+ f)) = 2\omega^{\mu\nu} \nabla_\mu (  \omega_\nu{}^\rho \nabla_\rho \log(\lambda_+ f)) = - 2 \nabla^\mu \nabla_\mu \log(\lambda_+ f).
\ee
This gives a simple formula for the scalar curvature of the K\"ahler metric, which we will need later. The reason why we wrote the function arising here as $\lambda_+ f$ will become clear later. 
	
For later use, we mention another important identity related to the K\"ahler connection $a$, and the conformal factor. We have
\begin{align}\label{a-df}
	(a,d\lambda_+) = (a,df)=0.
\end{align}
Both are true for exactly the same reason, so we only show the first one. This holds because
\begin{align}\label{a-d-lambda}
	(a,d\lambda_+) \sim \Sigma^1 \wedge d \log \left( \frac{CD}{(r + \im a q){}^4}\right)\wedge d\lambda_+ =0,
\end{align}
because of $\Sigma^1$ containing either $dr$ or $dq$, and thus vanishing when wedged with $dr\wedge dq$.

\section{K\"ahler version of the Teukolsky Master Equation}
\label{sec:teuk}

The purpose of this section is to show that the Teukolsky Master Equation for spin-k extremal weight scalars can be rewritten in terms of K\"ahler geometry, thus proving Theorem A of the Introduction. We will be using the identities obtained in Appendix C heavily. 

\subsection{Teukolsky Master Equation}

The homogeneous (no stress-energy tensor) Teukolsky spin $k$ equation from \cite{Teukolsky:1973ha} is
\begin{equation}\label{O-T}
\begin{aligned}
\mathcal{O}_k^T = &\Bigg[
\frac{(r^2 + a^2)^2}{\Delta} - a^2 \sin^2\theta
\Bigg]\partial_t^2 
+ \frac{4Mar}{\Delta}\partial_t \partial_\phi 
+ \Bigg[
\frac{a^2}{\Delta} - \frac{1}{\sin^2\theta}
\Bigg]\partial_\phi^2 
\\
&\quad
- \Delta^{-k}\partial_r \left( \Delta^{k+1}\partial_r \right)
- \frac{1}{\sin\theta}\partial_\theta \left( \sin\theta \partial_\theta \right)
\\
&- 2k \Bigg[ \frac{M(r^2 - a^2)}{\Delta}- r + i a \cos\theta \Bigg]\partial_t - 2k \Bigg[
\frac{a(r-M)}{\Delta} - \frac{i\cos\theta}{\sin^2\theta} \Bigg]\partial_\phi 
\\
&\quad
+ \left(k^2 \cot^2\theta - k\right).
\end{aligned}
\end{equation}

 \subsection{The K\"ahler operator}
 
 Our task is to show that Teukosly operator $\mathcal{O}_k^T$ is related by a similarity transformation to the following K\"ahler geometry operator
\begin{align}\label{K-operator}
\mathcal{O}_k \Phi^{(k)} &= (-(\nabla^\mu +\im k a^\mu)(\nabla_\mu + \im k a_\mu) + \frac{1+2k^2}{6} s)\Phi^{(k)} \\ \nonumber
&= -\nabla^\mu \nabla_\mu \Phi^{(k)} -2 \im k(a,d \Phi^{(k)}) + (k^2(a,a)+\frac{1+2k^2}{6}s) \Phi^{(k)}.
\end{align}

We have given the form of $a_\mu$ in (\ref{a-f}). It can be written as
\begin{align}
a_\mu = \omega_\mu{}^\nu \partial_\nu \log( \lambda_+ f)= \omega_\mu{}^\nu \partial_\nu( \log(f) + \log(\lambda_+)).
\end{align}
Thus, we have
\begin{align}
	(a,a) & = (d\log(f),d\log(f)) +  2(d\log(f),d\log(\lambda_+))+  (d\log(\lambda_+),d\log(\lambda_+)).
\end{align}

With the benefit of hindsight, we introduce the new fields 
\be\label{new-phi}
\phi^{(k)} = f \tp^{(k)}.
\ee 
We would like to rewrite the operators in (\ref{O-F}) as those on these fields. We have
\begin{align}\label{eq:apx-conformal-box-transform-1}
	\nabla^\mu \nabla_\mu (f^{-1} \phi^{(k)})  = f^{-1} \left[ \nabla^\mu \nabla_\mu \phi^{(k)} - 2(d\log(f),d\phi^{(k)})  +  (2( d\log(f),d\log(f))- f^{-1} \nabla^\mu \nabla_\mu f)   \phi^{(k)} \right].
\end{align}
Using (\ref{nabla-squared}) with (\ref{gamma}) we have
\begin{align}
	\nabla^\mu \nabla_\mu \phi^{(k)} -  2(d\log(f), d\phi^{(k)}) =  g^{\mu\nu} \partial_\mu \partial_\nu \phi^{(k)}.
	\end{align}
This means we get
\begin{align}
\mathcal{O}_k \Phi^{(k)} = f^{-1} \Big[ - g^{\mu\nu} \partial_\mu \partial_\nu \phi^{(k)}  
  - 2 \im k (a, d \phi^{(k)} ) + V  \phi^{(k)}    \Big],
  \end{align}
 where  
\begin{align}
V=  \frac{1+2k^2}{6} s + f^{-1} \nabla^\mu \nabla_\mu f   - 2(d\log(f),d\log(f)) + k^2(a,a).
\end{align}
If we substitute (\ref{delta-f-ident}) and (\ref{s}) we get
\begin{align}\nonumber
V &= -2(1+2k^2) (d\log(f),d\log(\lambda_+)) 
- ( d\log(f),d\log(f)) + 4(d\log(f),d\log(\lambda_+))- ( d\log(\lambda_+),d\log(\lambda_+))\\ \nonumber
&+ k^2  \left( (d\log(f),d \log(f))+2(d\log(f),d\log(\lambda_+))+  (d\log(\lambda_+),d \log(\lambda_+))\right) \\ \nonumber
&=(k^2 -1) \left(  (d\log(f),d \log(f))-2(d\log(f),d\log(\lambda_+))+  (d\log(\lambda_+),d \log(\lambda_+))\right) = (k^2-1)(\tilde{a},\tilde{a}).
\end{align}
We rewrote the final result in terms of the connection $\tilde{a}$ introduced in (\ref{tilde-a}). 

We now write down the operators $g^{\mu\nu} \partial_\mu\partial_\nu$ and $2\im k a^\mu \partial_\mu$ explicitly. As we observed in (\ref{inverse-metric}), the operator $\rho^2 g_{Kerr}^{-1}$ is separable, and the arising here operators are separable for the same reason. Indeed, using $\rho^2=(\lambda_+ \lambda_-)^{-1}$ and (\ref{inverse-metric}), we have
\be
g^{-1} = \lambda_+^{-2} g^{-1}_{Kerr} =\frac{\lambda_+^{-2}}{\rho^2} \rho^2 g^{-1}_{Kerr} = \frac{\lambda_-}{\lambda_+} ( g_1^{-1} + g_2^{-1}).
\ee
Using (\ref{g1-g2}) and (\ref{a-dF}) we have
\begin{align}
	\frac{\lambda_+}{\lambda_-}g^{\mu\nu} \partial_\mu \partial_\nu  & = -\frac{1}{C}((r^2+a^2)\partial_t + a \partial_\phi)^2  + C \partial_r^2  + \frac{1}{D} (a(1-q^2) \partial_t + \partial_\phi)^2  + D \partial_q^2 , \\
	\frac{\lambda_+}{\lambda_-} 2 \im k a^\mu \partial_\mu & = -4k(r-\im a q) \partial_t  + k \frac{\partial_r C}{C} ((r^2+a^2) \partial_t + a \partial_\phi)  + k\frac{\im \partial_q D}{D} ( a(1-q^2) \partial_t + \partial_\phi) .
\end{align}
This gives
\begin{align}\label{O-k-prime}
	 \mathcal{O}_k = \frac{\lambda_-}{\lambda_+} f^{-1} \mathcal{O}'_k f,
	\end{align}
where 
\begin{align}	
	\mathcal{O}'_k= \frac{1}{C}((r^2+a^2)\partial_t + a \partial_\phi)^2  - C \partial_r^2  + 4k r \partial_t  -k \frac{\partial_r C}{C} ((r^2+a^2) \partial_t + a \partial_\phi) 
	 \nonumber\\ - \frac{1}{D} (a(1-q^2) \partial_t + \partial_\phi)^2  - D \partial_q^2 -k \left(4\im a q \partial_t  + \frac{\im \partial_q D}{D} ( a(1-q^2) \partial_t + \partial_\phi)  \right) +V',
\end{align}
and
\begin{align}
V' = (k^2-1)(g_1^{-1}(\tilde{a},\tilde{a}) + g_2^{-1}(\tilde{a},\tilde{a})) =(k^2-1) \left( \cot^2\theta + \frac{(r-M)^2}{\Delta}\right),
\end{align}
where we used (\ref{t-a-squared}) two write the last equality. It is clear that this operator separates in its dependence on $r$ and $q$. In particular, the potential $V'$ separates.

\subsection{Similarity transformation}

We first rewrite the operator $\mathcal{O}'_k$ we obtained in a more standard notations where $C=\Delta= r^2-2Mr +a^2$. We also use $q=\cos\theta$ and $D=1-q^2=\sin^2\theta$. We get
\begin{align}	\label{Os}
	\mathcal{O}'_k= \Bigg[\frac{(r^2 + a^2)^2}{\Delta} - a^2 \sin^2\theta \Bigg]\partial^2_t + \frac{4M ar}{\Delta}\partial_t \partial_\phi 
	+\Bigg[\frac{a^2}{\Delta} - \frac{1}{\sin^2\theta} \Bigg] \partial^2_\phi
	\nonumber\\   
	   - \Delta \partial_r^2  - \sin\theta \left( \partial_\theta \frac{1}{\sin\theta} \partial_\theta \right)
	\nonumber\\ 		
		- 2k\Bigg[ \frac{M(r^2 - a^2)}{\Delta}- r + \im a \cos\theta\Bigg] \partial_t 
		- 2k\Bigg[ \frac{a(r-M)}{\Delta}   - \frac{\im \cos\theta}{\sin^2\theta} \Bigg] \partial_\phi 
		\nonumber\\ 		
		+(k^2-1) \left( \cot^2\theta + \frac{(r-M)^2}{\Delta}\right).
\end{align}
Comparing this to the Teukolsky operator (\ref{O-T}) we see that the first and third lines in these operators match exactly, also the second order parts of the second lines match. This suggests that the two operators are related by a conjugation by a function of $r,q$. 

Such a conjugation is easily found. First, we observe that 
\be
\Delta^{-m} (  \Delta^k \partial_r (\Delta^{k+1} \partial_r (\Delta^m \phi))) =  \Delta \partial_r^2 \phi + (2m+k+1) \partial_r \Delta \partial_r \phi + \left( m \partial_r^2 \Delta + m(m+k) \frac{(\partial_r\Delta)^2}{\Delta}\right)\phi.
\ee
There is no first derivative term here if
\be
m = - \frac{1+k}{2}.
\ee
In this case the operator simplifies to
\be
\Delta^{-m} (  \Delta^{-k} \partial_r (\Delta^{k+1} \partial_r (\Delta^m \phi))) =  \Delta \partial_r^2 \phi - \left((k+1) + \frac{(k^2-1)(r-M)^2}{\Delta}\right) \phi,
\ee
which becomes particularly simple when $k=\pm 1$. 

Second, we consider 
\be
\frac{1}{\sin^{n+1}\theta}\partial_\theta \left( \sin\theta \partial_\theta (\sin\theta^n \phi )\right)=\partial_\theta^2 \phi + (2n+1) \frac{\cos\theta}{\sin\theta} \partial_\theta \phi + \left( n^2 \frac{\cos^2\theta}{\sin^2\theta} -n \right) \phi.
\ee
The first two terms here match the operator in $\mathcal{O}'_k$
\be
 \sin\theta \left( \partial_\theta \frac{1}{\sin\theta} \partial_\theta \phi \right) = \partial_\theta^2 \phi - \frac{\cos\theta}{\sin\theta} \partial_\theta \phi
 \ee
 if $n=-1$. Thus
 \be
 \partial_\theta \left( \sin\theta \partial_\theta (\frac{1}{\sin\theta} \phi )\right) =  \sin\theta \left( \partial_\theta \frac{1}{\sin\theta} \partial_\theta \phi \right) + \left(  \cot^2\theta +1 \right) \phi.
 \ee
 It is now clear that the two operators match after a conjugation
 \be
 (\sin\theta \Delta^{(1+k)/2}) \mathcal{O}_k^T  (\sin\theta \Delta^{(1+k)/2})^{-1} = \mathcal{O}'_k.
 \ee
 Combining with (\ref{O-k-prime}), and recalling that $f= \sqrt{\Delta} \sin\theta \lambda_+$, we finally get the formula (\ref{comparison})
 \begin{align}
 \mathcal{O}^T_k = \frac{\lambda_+}{\lambda_-} \frac{\lambda_+}{\Delta^{k/2}} \mathcal{O}_k \left( \frac{\Delta^{k/2}}{\lambda_+} \right).
 \end{align}

\section{Spin one Teukolsky equation on Kerr}
\label{sec:spin-one}

In this section we interpret the spin-one Teukolsky equation as the operator $d_+ d_+^*$, where $d_+: \Lambda^1 \to \Lambda^+$ is the truncated Hodge operator and $d_+^*$ is its adjoint. The operator $d_+$ is conformally invariant, while $d_+^*$ depends on the metric. The separable spin one Teukolsky operator is $d_+ d_+^*$ for the (complex) K\"ahler metric. Our main aim is to show how the operator $\mathcal{O}_k$, with $k=\pm 1$, introduced in (\ref{K-operator}) arises from $d_+ d_+^*$. 

\subsection{Conformal invariance of Maxwell's equations in four dimensions}

We will use formalism of differential forms, in which Maxwell's equations for a connection 1-form $\mathcal{A} \in \Lambda^1$ take the form
\begin{align}
	\F = d\mathcal{A}, \quad \delta \F = 0.
\end{align}
Here $\delta$ is the adjoint of the exterior derivative operator $d$, and is defined by $\delta = -\star d \star$, where $\star$ denotes the Hodge dual. 

In 4 dimensions we have
\begin{align}
	\frac{1}{2} \star d \F +  \frac{i}{2} \delta \F = \star d \frac{1}{2} (1 -i \star) \F = \star d \F_+ = i \delta \F_+.
\end{align}
Here $\F_+$ is the self-dual part of the curvature 2-form. This means that the Maxwell's equation $\delta \F=0$ is equivalent to the chiral equation $\delta \F_+=0$. This means that we can alternatively encode Maxwell's equations in the following chiral form
\begin{align}
	\F_+ = \frac{1}{2}(1-i\star) d \mathcal{A}, \quad \delta \F_+ = 0.
\end{align}

Another very useful way of writing these equations is to introduce
\begin{align}
d_+ = P_+ d = \frac{1}{2}(1-i\star)d, \quad d_+^* = \delta P_+ = - \star d \star P_+ = -\im \star d P_+.
\end{align}
Then $\F_+= d_+ \mathcal{A}$, and the chiral version of the Maxwell's equations is $d_+^* d_+ \mathcal{A}=0$ or 
\begin{align}
	d_+^* \F_+ = - \star d \star P_+ \F_+ =  - \im \star d  \F_+ = - \im \star d \frac{1}{2} (1 - \im \star) \F = - \frac{\im}{2} \star d \F + \frac{1}{2} \delta \F = 0.
\end{align}
The real and imaginary parts of this equation are then $d\F = 0$ and $\delta \F = 0$ respectively. 

The main purpose of the proceeding discussion is to make clear the well-known fact that Maxwell's equations, and in particular Maxwell's equations in the chiral form, are conformally invariant. Indeed, we have written them in the chiral form $d_+^* d_+ \mathcal{A}=0$. The operators $d_+= P_+ d$ is constructed solely from the operator of the exterior derivative, which is metric independent, and the Hodge star operator on 2-forms, which is conformally invariant. Thus $d_+$ is conformally invariant. The operator $d_+^* =- \star d \star P_+$. All operators here, except the last Hodge star, are conformally invariant. The last Hodge operator is on 3-forms, and is not invariant, but still transforms under conformal transformations by a simple rescaling. This implies that if  $d_+^* d_+ \mathcal{A}=0$ holds in one metric, it also holds in a conformally related metric. We can use this fact to work with Maxwell's equations on the background of the K\"ahler metric $g$ instead. 

\subsection{Teukolsky spin-one Operator as the Hodge Laplacian on $\Lambda^+$}

Everywhere in this section all operators are with respect to the K\"ahler metric $g$. On the background of the K\"ahler metric $g$, which is conformal to the Kerr metric, Maxwell's equations become $d_+^* \F= d_+^* \F_+=0$. Thus, they are first-order differential equations on the self-dual part $\F_+=P_+ \F$ of the field strength $\F=d\mathcal{A}$. The object $\F_+$ can be decomposed into the basis of self-dual 2-forms $\Omega^i$
\begin{align}
	\F_+ = \Phi^i \Omega^i.
\end{align}
The equations $d_+^* \F_+=0$ is then a set of coupled equations on the scalars $\Phi^i$. Let us rewrite these equations using the more standard in the literature notations as
\begin{align}
	\mathcal{E} \F_+ =0, \qquad	\mathcal{E} : \Lambda^+ \to \Lambda^1, \qquad \mathcal{E} = d_+^*.
\end{align}
Teukolsky applies to these equations the decoupling operator $\mathcal{S}$, which is a first-order differential operator chosen so that $\mathcal{S} \mathcal{E} = \mathcal{O}$ is a second-order differential operator acting on the self-dual scalars $\Phi^i$ in such a way that these three scalars decouple. Given that we are looking for an operator $\mathcal{O}: \Lambda^+ \to \Lambda^+$, and $\mathcal{E}: \Lambda^+ \to \Lambda^1$, our decoupling operator $\mathcal{S}$ must be a map 
\begin{align}
	\mathcal{S} : \Lambda^1 \to \Lambda^+. 
\end{align}
There already is a natural such operator, which is $d_+$. This suggests we identify 
\begin{align}
	\cS = 2d_+,
\end{align}
where the numerical factor is for agreement with previous formulas, and consider the operator 
\begin{align}
	\mathcal{O}= 2d_+ d_+^*.
\end{align}
Note that our identity $\mathcal{S} \mathcal{E} = \mathcal{O}$ is not in conflict with the more standard Teukolsky-Wald identity $\mathcal{S} \mathcal{E} = \mathcal{O}\mathcal{T}$ because $\mathcal{E}$ acts on the field strength and not on the gauge potential. If we considered the gauge potential $\mathcal{A}$ as the main variable instead, then $\mathcal{E}$ changes to $\mathcal{E} = d_+^* d_+$ and the usual Teukolsky-Wald relation $\mathcal{S} \mathcal{E} = \mathcal{O}\mathcal{T}$ is reproduced with $\mathcal{T}=d_+$. 
 
We now manipulate this operator using standard differential geometric identities. First, we have
\begin{align}
	d_+ d^*_+ =  P_+ d \delta P_+.
\end{align}
Now, by exploiting that $ \star P_+ = P_+ \star = \im P_+$ and $\star^2 = -1$ on 2-forms it is easy to show that 
\begin{align}
	P_+ d \delta P_+ = P_+ \delta d P_+ = -\im P_+ d \star d P_+.
\end{align}
Recalling the definition of the Hodge Laplace operator on forms
\begin{align}
\Delta_H = d\delta + \delta d,
\end{align}
we see that 
\begin{align}
	2d_+ d^*_+ =  P_+ \Delta_H P_+.
\end{align}
But Hodge Laplacian commutes with the Hodge star $\Delta_H \star = \star \Delta_H$, and thus $\Delta_H$ commutes with $P_+$. Thus, we can also write
\begin{align}
	\mathcal{O} =  \Delta_H P_+.
\end{align}
Thus, in words, up to the factor of two, the operator $\mathcal{O}=2d_+ d^*_+$ is the restriction of the Hodge Laplacian to the space of self-dual 2-forms $\Lambda^+$ for the K\"ahler metric $g$. The main spin-one claim of this paper is that this operator is the separable spin-one Teukolsky operator. We will verify this by computing this operator explicitly. 

\subsection{Weizenbock formula}

Let us now use the standard Weitzenbock formula, see e.g. \cite{Brun-Weitz}, which states the Hodge Laplacian on 2-forms can be rewritten in terms of the rough Laplacian and the curvature terms
\begin{align}
	\Delta_H B = \nabla^* \nabla  B + \frac{s}{3} B - 2 W(B).
\end{align}
Here $B\in \Lambda^2$, $\nabla^* \nabla = - g^{\mu\nu} \nabla_\mu \nabla_\nu$ is the rough Laplacian, $s$ is the scalar curvature $s=R_{\mu\nu}g^{\mu\nu}$, and $W(B){}_{\mu\nu} = (1/2) W_{\mu\nu}{}^{\rho\sigma} B_{\rho\sigma}$ with $W_{\mu\nu\rho\sigma}$ being the Weyl tensor. We give a proof of this formula in the Appendix. 

We can apply the Weitzenbock formula to $\F_+$. This gives 
\begin{align}\label{O-weitz}
	2d_+ d_+^* ( \tp^i \Omega^i) = \nabla^* \nabla (\tp^i \Omega^i) + \frac{s}{3}  \tp^i \Omega^i - 2 \tp^i W(\Omega){}^i.
\end{align}

Up to this point, there was nothing about our discussion that was specific to the metric being K\"ahler. The formula (\ref{O-weitz}) is valid for any metric. We now specialise to the case of a K\"ahler metric.

Let us rewrite the decomposition $\F_+=\tp^i \Omega^i$ into the self-dual curvature scalars in the basis of $\Omega^1=\omega, \Omega^\pm$
\begin{align}
\F_+ = \Phi \omega + \Phi^- \Omega^+ + \Phi^+ \Omega^-.
\end{align}
Then, using the form (\ref{weyl-kahler}) of the self-dual Weyl curvature in a K\"ahler metric we have
\begin{align}
	&\frac{s}{3}  \tp^i \Omega^i - 2 \tp^i  W(\Omega){}^i = \frac{s}{3}  ( \Phi \omega + \Phi^- \Omega^+ + \Phi^+ \Omega^-) - 2 \frac{s}{6} ( \Phi \omega -\frac{1}{2} \Phi^- \Omega^+ -\frac{1}{2} \Phi^+ \Omega^-)= 
	&\frac{s}{2}( \Phi^- \Omega^+ + \Phi^+ \Omega^-).
\end{align}
In other words, only the terms proportional to $\Omega^\pm$ arise in the curvature terms. 

The next step is to rewrite the rough Laplacian terms in (\ref{O-weitz}). To do this, we use the fact that the K\"ahler covariant derivatives of the self-dual 2-forms $\omega,\Omega^\pm$ satisfy (\ref{nabla-omega-Omega}). Using these identities we can shift the covariant derivatives by terms involving $a_\mu$, after which the self-dual 2-forms $\Omega^\pm$ become covariantly constant and can be pulled out from under the derivative operators. This gives
\begin{align}\label{O-F}
	\mathcal{O} \F_+  
		 = -\left( \nabla^\mu \nabla_\mu \tp \right) \omega - \left( (\nabla^\mu - \im a^\mu)(\nabla_\mu - \im a_\mu) \tp^- - \frac{s}{2} \tp^-\right) \Omega^+ -\left( (\nabla^\mu + \im a^\mu)(\nabla_\mu + \im a_\mu) \tp^+ - \frac{s}{2} \tp^+\right) \Omega^-.
\end{align}
The operators $\mathcal{O}_{\pm 1}$ are showing their appearance here. This proves Theorem B of the Introduction. From the results of the previous section we know that they are related to the spin-one Teukolsky operator by a similarity transformation. For the scalar $\Phi$ we recover the Fackerell-Ipser equation \cite{Fackerell:1972hg}, which is not separable unless the spin parameter is set to zero (i.e. $a = 0$). 

\subsection{Alternative computation}

We now recompute the operator $d_+ d_+^*$ in an alternative fashion, which will be useful for the spin two considerations in the next subsection. The operator $d_+: \Lambda^1 \to \Lambda^+$. We will parametrise the space $\Lambda^+$ by projecting the self-dual 2-forms onto the basis of 2-forms given by $\Omega^i$. Thus, we will describe sections of $\Lambda^+$ as fields of 3-vectors $\Phi^i$. Then 
\[
d_+: \Lambda^1 \ni \xi_\mu \to \Phi^i= \frac{1}{2} \Omega^{i\mu\nu} \nabla^a_\mu \xi_\nu \in \Lambda^+
\]
The adjoint operator is 
\[
d_+^*: \Phi^i \to \xi_\mu = \Omega^i_\mu{}^\nu \nabla^a_\nu \Phi^i.
\]
Here $\nabla^a_\mu$ is the covariant derivative that acts on all indices of a tensor, including the internal index $i$. In the case of $\Phi^i$ that has only the internal index $\nabla^a_\mu \Phi^i = \partial_\mu \Phi^i + \epsilon^{ijk} a^j_\mu \Phi^k$. We will also be using the fact that 
\be\label{cov-const}
\nabla^a_\mu \Omega^i_{\rho\sigma} = 0.
\ee

The composition of the two operators is
\begin{align}
2(d_+ d_+^* \Phi)^i &=  \Omega^{i\mu\nu} \nabla^a_\mu \Omega^j_\nu{}^\alpha \nabla^a_\alpha \Phi^j = \Omega^{i\mu\nu}  \Omega^j_\nu{}^\alpha \nabla^a_\mu \nabla^a_\alpha \Phi^j =
(-\delta^{ij} g^{\mu\alpha} + \epsilon^{ijk} \Omega^{k\mu\alpha}) \nabla^a_\mu \nabla^a_\alpha \Phi^j \\ \nonumber
&= - \nabla^{a\mu} \nabla^a_\mu \Phi^i + \frac{1}{2}  \epsilon^{ijk} \Omega^{k\mu\alpha} \epsilon^{jmn} F^m_{\mu\alpha} \Phi^n.
\end{align}
To get the second equality we have used (\ref{cov-const}), and to get the third expression we have used the algebra (\ref{omega-algebra}) of 2-forms $\Omega^i_{\mu\nu}$. The last expression is obtained using $\nabla^a_{[\mu} \nabla^a_{\nu]} \Phi^i = (1/2) \epsilon^{ijk} F^j_{\mu\nu} \Phi^k$. We now use
\be
F^i_{\mu\nu} = M^{ij} \Omega^j_{\mu\nu} + \text{ASD terms}.
\ee
This gives
\[
\Omega^{i\mu\nu} F^j_{\mu\nu} = 4 M^{ij}, \qquad M^{ij} = \frac{s}{4} L^i L^j.
\]
Using this, we get
\begin{align}
2(d_+ d_+^* \Phi)^i &=- \nabla^{a\mu} \nabla^a_\mu \Phi^i + \frac{s}{2} \epsilon^{ijk}  \epsilon^{jmn} L^k L^m \Phi^n.
\end{align}
We now decompose 
\[
\Phi^i = \Phi L^i + \Phi^+ \bar{S}^i + \Phi^- S^i.
\]
Then, using (\ref{vector-product}), we have
\[
 \epsilon^{jmn} L^m \Phi^n =\im \Phi^+ \bar{S}^k  - \im \Phi^- S^k, 
 \]
and then
\[
 \epsilon^{ijk}  \epsilon^{jmn} L^k L^m \Phi^n = \Phi^+ \bar{S}^i  + \Phi^- S^i.
 \]
 This gives the final result
 \be
 2(d_+ d_+^* \Phi)^i =- \nabla^{a\mu} \nabla^a_\mu \Phi^i + \frac{s}{2} ( \Phi^+ \bar{S}^i  + \Phi^- S^i).
 \ee
 This matches what we have obtained using the Weitzenbock formula. 
 
 \subsection{Spin two attempt}
 
The main purpose of this subsection is to present a natural generalisation of the spin-one operator $d_+ d_+^*$ to the spin-two case. As is clear from the discussion that follows, this generalisation, while realising some of the expectations in the spin-two case, is nevertheless inadequate, because the full separable Teukolsky operator is not reproduced. Nevertheless, we include the following calculation into this paper because it suggests the natural direction of generalisation from spin-one to spin-two. Our belief is that some modification of the construction below should work, and this is the reason for describing it.
 
Thus, motivated by the story in the spin-one case, the most natural guess for the spin-two generalisation is to consider
the operator $d^{(2)}_+: \Lambda^1\otimes \Lambda^+ \to S^2_0( \Lambda^+)$ given by
\be\label{d-plus-spin-two}
d^{(2)}_+: \xi^i_\mu \to \psi^{ij} = P^{ij|kl} \Omega^{k\mu\nu} \nabla^a_\mu \xi^{l}_\nu, \qquad P^{ij|kl} = \frac{1}{2} ( \delta^{ik} \delta^{jl} + \delta^{il}\delta^{jk}) - \frac{1}{3} \delta^{ij} \delta^{kl}.
\ee
Here $P^{ij|kl}$ is the projector onto the symmetric tracefree part. Its adjoint is given by
\be
(d^{(2)}_+)^* : \psi^{ij} \to \xi^i_\mu = \Omega_\mu^j{}^\nu \nabla^a_\nu \psi^{ij}.
\ee

The composition is given by
\begin{align}\nonumber
(d^{(2)}_+ (d^{(2)}_+)^* \psi)^{ij} &=P^{ij|kl} \Omega^{k\mu\nu} \nabla^a_\mu  \Omega_\nu^n{}^\alpha \nabla^a_\alpha \psi^{ln} = P^{ij|kl} \Omega^{k\mu\nu}  \Omega_\nu^n{}^\alpha  \nabla^a_\mu \nabla^a_\alpha \psi^{ln} = P^{ij|kl} (-\delta^{kn} g^{\mu\alpha} + \epsilon^{knm} \Omega^{m\mu\alpha} ) \nabla^a_\mu \nabla^a_\alpha \psi^{ln}
\\ 
&= P^{ij|kl}\left( - \nabla^{a\mu} \nabla^a_\mu \psi^{kl} + \frac{1}{2} \epsilon^{knm} \Omega^{m\mu\alpha} ( \epsilon^{lpq} F^p_{\mu\alpha} \psi^{qn} +  \epsilon^{npq} F^p_{\mu\alpha} \psi^{lq})\right)\\ \nonumber
&= P^{ij|kl}\left( - \nabla^{a\mu} \nabla^a_\mu \psi^{kl} + \frac{s}{2} \epsilon^{knm} ( \epsilon^{lpq} L^m L^p \psi^{qn} +  \epsilon^{npq} L^m L^p \psi^{lq})\right),
\end{align} 
where we have used $F^i_{\mu\nu} = M^{ij} \Omega^j_{\mu\nu}$ and $M^{ij} = (s/4) L^i L^j$. We then decompose
\be
\psi^{ij} = \Psi^+ \bar{S}^i \bar{S}^j + \Psi^{L+} L^{(i} \bar{S}^{j)} + \Psi ( 2L^i L^j - S^i\bar{S}^j - \bar{S}^i S^j) + \Psi^{L-} L^{(i} S^{j)} + \Psi^- S^i S^j.
\ee
We have
\be
\epsilon^{lpq} L^p \psi^{qn} = \im \left( \Psi^+ \bar{S}^l \bar{S}^n + \frac{1}{2} \Psi^{+L} \bar{S}^l L^n + \Psi (S^l \bar{S}^n - \bar{S}^l S^n) - \frac{1}{2} \Psi^{L-} S^l L^n - \Psi^- S^l S^n\right),
\ee
and so
\begin{align}
&\frac{s}{2} \epsilon^{knm} ( \epsilon^{lpq} L^m L^p\psi^{qn} +  \epsilon^{npq} L^m L^p \psi^{lq})= - s  \epsilon^{kmn}  L^m \epsilon^{(l|pq} L^p\psi^{q|n)} \\ \nonumber
&= s  \left( \Psi^+ \bar{S}^l \bar{S}^k + \frac{1}{4} \Psi^{+L} \bar{S}^l L^k  + \frac{1}{4} \Psi^{L-} S^l L^k + \Psi^- S^l S^k\right).
\end{align}
Thus, finally
\begin{align}
(d^{(2)}_+ (d_+^{(2)})^* \psi)^{ij} = - \nabla^{a\mu} \nabla^a_\mu \psi^{ij} + s  \left( \Psi^+ \bar{S}^i \bar{S}^j + \frac{1}{4} \Psi^{+L} \bar{S}^{(i} L^{j)}  + \frac{1}{4} \Psi^{L-} S^{(i} L^{j)} + \Psi^- S^i S^j\right).
\end{align}
The coefficient in front of $s$ for the extremal scalars is unity rather than the desired $3/2$, which the formula (\ref{K-operator}) requires when $k=\pm 2$. So, this is not the operator that gives the separable Teukolsky spin-two equations. Nevertheless, the derivative part of the arising operator on the extreme weight scalars $\Psi^\pm$, namely 
\[
\nabla^{a\mu} \nabla^a_\mu = (\nabla^\mu \pm 2\im a^\mu)(\nabla_\mu \pm 2\im a_\mu)
\]
is precisely the one in the operator (\ref{K-operator}). So, this approach appears to be in the right direction, and we believe that some modification of the operators $d^{(2)}_+, (d_+^{(2)})^*$ can produce the correct potential term as well. 

\section{Discussion}

In this paper we have shown that the decoupling of Kerr perturbations has a direct geometric origin: it arises from the K\"ahler structure to which the Kerr metric is conformal. In particular, we demonstrated that the spin-k Teukolsky equation has a simple interpretation (\ref{O-k-intr}) in terms of K\"ahler geometry, and that in K\"ahler geometry decoupling follows from the parallel decomposition of self-dual 2-forms. We have also shown that the spin-one Teukolsky operator is essentially the Hodge Laplacian operator in K\"ahler geometry. At a structural level, the mechanism can be traced to the reduction of the self-dual part of the Levi-Civita connection from a (complexified in the Lorentzian signature) ${\rm SU}(2)$ connection to a (complexified) ${\rm U}(1)$ connection, which ensures that the natural differential operators preserve the decomposition of $\Lambda^+$ into one-dimensional sub-bundles and hence do not mix different spin weight components.

Although our analysis was carried out in the Kerr setting, the underlying mechanism is not specific to this case. It applies more generally to the Pleba\'nski–Demia\'nski family of solutions \cite{Plebanski:1976gy}, for which the metric is conformally related to a Kähler geometry with analogous properties. Thus, the geometric origin of decoupling identified here should be viewed as a structural feature of this entire class of type D spacetimes, rather than a special property of Kerr.

Our results, while clearly showing that K\"ahler geometry underlies decoupling of Kerr perturbations, are incomplete in two directions. First, our preliminary spin-two analysis in the previous section did not produce the desired Teukolsky operator for the extreme weight scalars. The Laplace-type operator $d^{(2)}_+ (d_+^{(2)})^*$ produced the correct derivative part of the operator $\mathcal{O}_{\pm 2}$, but did not give the right coefficient in front of the the scalar curvature potential term. This mismatch is not accidental. The operator $d^{(2)}_+ (d_+^{(2)})^*$ decouples all five components of the spin-two field, whereas existing results \cite{Aksteiner:2010rh}, \cite{Araneda:2016iwr} indicate that such a complete decoupling is incompatible with the structure of the linearised Einstein equations without gauge fixing. In particular, the spin-two system necessarily involves gauge-dependent components, and decoupling of intermediate spin weights requires a delicate interplay between geometry and gauge conditions. So, $d^{(2)}_+ (d_+^{(2)})^* \psi^{ij}=0$ is not the correct equation describing the linearisation of the Weyl curvature on the Kerr background. And indeed, unlike the spin-one case where we were able to derive the equation $d_+ d_+^* \mathcal{F}=0$ from Maxwell's equations satisfied by the field strength $\mathcal{F}$, no such link between $d^{(2)}_+ (d_+^{(2)})^* \psi^{ij}=0$ and linearised Einstein equations on Kerr was established. Nevertheless, the Pleba\'nski formalism suggests that there is such a link, but it is much more involved than that in the linear Maxwell case. We hope to come back to this in a separate publication. 

The second point we did not treat in this paper is reconstruction. As explained in \cite{Green}, the differential-forms approach also bears on the construction of Lorenz-gauge electromagnetic potentials; a dedicated treatment will be given elsewhere. We believe that an analogous reconstruction in the spin-two case will require going beyond the usual Hodge decomposition theory, to at least bundle-valued differential forms as in (\ref{d-plus-spin-two}).

In summary, our results show that K\"ahler  geometry provides the underlying reason for the decoupling of Kerr perturbations, as demonstrated at the level of Teukolsky operator by Theorem A of the Introduction for arbitrary spin, and at the detailed field equations level for spin-one fields. In the spin-two case, the difficulty is in the derivation of a closed second-order equation for the linearised Weyl tensor directly from the linearised Einstein equations. In general, the resulting equation involves not only the linearised Weyl tensor, but also the metric perturbation, and is therefore not closed. The central challenge is to identify a gauge in which the system reduces to equations involving only the Weyl tensor. Important progress in this direction has been made in \cite{Aksteiner:2010rh, Araneda:2016iwr}, but these approaches rely on spinorial methods, which tend to obscure the underlying geometry. The results of the present paper suggest that a formulation based on Pleba\'nski variables and K\"ahler geometry should provide a more direct geometric understanding, with decoupling built in. Developing such a formulation remains the problem for future work.

\appendix

\section{Proof of Weizenbock formula}

We start by proving
\be
(\Delta_H B)_{\mu\nu} = (\nabla^* \nabla B)_{\mu\nu} + R_\mu{}^{\alpha} B_{\alpha\nu} - R_\nu{}^{\alpha} B_{\alpha\mu} -  R_{\mu\nu}{}^{\alpha\beta} B_{\alpha\beta}.
\ee
Here
\be
\Delta_H = d \delta + \delta d,
\ee
with conventions
\be
(\delta B)_\nu = - \nabla^\mu B_{\mu\nu}, \qquad (\nabla^*\nabla B)_{\mu\nu} = - \nabla^\alpha \nabla_\alpha B_{\mu\nu}, \qquad
[\nabla_\mu,\nabla_\nu] \theta_\rho = R_{\mu\nu\rho}{}^\sigma \theta_\sigma.
\ee
Using this we have
\be
(\delta d B)_{\mu\nu} = - \nabla^\rho \nabla_\rho B_{\mu\nu} - \nabla^\rho \nabla_\mu B_{\nu\rho} - \nabla^\rho \nabla_\nu B_{\rho\mu},
\ee
and
\be
(d \delta B)_{\mu\nu} = \nabla_\mu (\delta B)_\nu - \nabla_\nu (\delta B)_\mu = - \nabla_\mu \nabla^\rho B_{\rho\nu} + \nabla_\nu \nabla^\rho B_{\rho\mu}.
\ee
This gives
\be
(\Delta_H B)_{\mu\nu} =- \nabla^\rho \nabla_\rho B_{\mu\nu} - [\nabla^\rho, \nabla_\mu] B_{\nu\rho} - [\nabla^\rho, \nabla_\nu] B_{\rho\mu}.
\ee
The commutator terms here are
\be
[\nabla^\rho, \nabla_\mu] B_{\nu\rho} = R^\rho{}_{\mu\nu}{}^\alpha B_{\alpha\rho} + R^\rho{}_{\mu\rho}{}^\alpha B_{\nu\alpha} = - R_{\mu}{}^\alpha B_{\alpha\nu} + R_{\mu}{}^\alpha{}_{\nu}{}^{\beta} B_{\alpha\beta}.
\ee
We can now use the first Bianchi identity
\[
R_{\mu}{}^\alpha{}_{\nu}{}^\beta + R^\alpha{}_{\nu\mu}{}^\beta + R_{\nu\mu}{}^{\alpha\beta}=0
\]
to get
\be\label{first-Bianchi}
R_{\mu}{}^\alpha{}_{\nu}{}^{\beta} B_{\alpha\beta} - R_{\nu}{}^\alpha{}_{\mu}{}^{\beta} B_{\alpha\beta} =R_{\mu\nu}{}^{\alpha\beta} B_{\alpha\beta} ,
\ee
so that
\begin{align}\label{hodge-2-forms}
(\Delta_H B)_{\mu\nu} =- \nabla^\rho \nabla_\rho B_{\mu\nu} + R_{\mu}{}^\alpha B_{\alpha\nu} - R_{\nu}{}^\alpha B_{\alpha\mu} - R_{\mu\nu}{}^{\alpha\beta} B_{\alpha\beta}.
\end{align}

In dimension four the Riemann curvature tensor decomposes as
\be\label{riemann-decomp}
R_{\mu\nu\rho\sigma} = W_{\mu\nu\rho\sigma} + \frac{1}{2} ( g_{\mu\rho} R_{\nu\sigma} - g_{\mu\sigma} R_{\nu\rho} - g_{\nu\rho} R_{\mu\sigma} + g_{\nu\sigma} R_{\mu\rho}) - \frac{s}{6} ( g_{\mu\rho} g_{\nu\sigma} - g_{\nu\rho} g_{\mu\sigma} ).
\ee
This means that
\be
R_{\mu\nu}{}^{\rho\sigma} B_{\rho\sigma} = W_{\mu\nu}{}^{\rho\sigma} B_{\rho\sigma} + R_{\mu}{}^\alpha B_{\alpha\nu}  - R_{\nu}{}^\alpha B_{\alpha\mu} - \frac{s}{3} B_{\mu\nu},
\ee
and so
\be
R_\mu{}^{\alpha} B_{\alpha\nu} - R_\nu{}^{\alpha} B_{\alpha\mu} -  R_{\mu\nu}{}^{\alpha\beta} B_{\alpha\beta} = \frac{s}{3} B_{\mu\nu} - W_{\mu\nu}{}^{\rho\sigma} B_{\rho\sigma}. 
\ee
This gives the final formula
\be
(\Delta_H B)_{\mu\nu} = - \nabla^\rho \nabla_\rho B_{\mu\nu} +  \frac{s}{3} B_{\mu\nu} - W_{\mu\nu}{}^{\rho\sigma} B_{\rho\sigma}.
\ee

\section{Null version of the \pleb{} formulation}

In the reviewed in section \ref{sec:kahler} K\"ahler geometry the basic objects were the 2-forms $\omega, \Omega^\pm$. The metric is encoded by these 2-forms, which span the space of self-dual 2-forms. The 2-form $\omega$ is the real K\"ahler form, while $\Omega^\pm$ are complex conjugates of each other. K\"ahler geometry is about Riemannian signature metrics, but there is an analogous description of the Lorentzian geometry, which we now develop. This gives us a version of the \pleb{} formalism, where instead of working with $\Sigma^{1,2,3}$ one introduces appropriate complex linear combinations of $\Sigma^{2,3}$. The main difference with the Riemannian signature case is that the Lorentzian $\Sigma^{1,2,3}$ are complex-valued, which results in $\Sigma^2\pm \im \Sigma^3$ to not be complex conjugates of each other. An alternative viewpoint is to say that we develop the null description of the \pleb{} formulation of general relativity. This description emulates the Newman-Penrose formalism, but this time for self-dual 2-forms rather than for the frame. The obtained formalism will be particularly useful in algebraically special spacetimes, in our case type D spacetimes.

The starting point is the same as in Newman-Penrose formalism. We introduce a coframe or equivalently a basis $e^0,e^1,e^2,e^3 \in \Lambda^1$ such that the Lorentzian spacetime metric is given by
\begin{align}
	g = - e^0 \otimes e^0 + e^1 \otimes e^1 + e^2 \otimes e^2 + e^3 \otimes e^3
\end{align}
The self-dual 2-forms corresponding to this choice of frame are
\begin{align}
	\Sigma^1 = \im e^0 \wedge e^1 - e^2 \wedge e^3, \quad \Sigma^2 = \im e^0 \wedge e^2 - e^3 \wedge e^1, \quad \Sigma^3 = \im e^0 \wedge e^3 - e^1 \wedge e^2.
\end{align}
To move to a null basis we introduce the following null frame
\begin{align}
	l = \frac{1}{\sqrt{2}}(e^0 - e^1),\quad n = \frac{1}{\sqrt{2}}(e^0 + e^1),\quad m = \frac{1}{\sqrt{2}}(e^2 + \im e^3),\quad \bar{m} = \frac{1}{\sqrt{2}}(e^2 - \im e^3).
\end{align}
The only non-zero inner products of the null basis are
\begin{align}
	g^{-1}(l,n) = -1, \quad g^{-1}(m, \bar{m}) = 1.
\end{align}
A similar null basis can be introduce for the self-dual 2-forms,
\begin{align}\label{sigmas-null-tetrad}
	\Sigma^1 = \im ( l \wedge n - m \wedge \bar{m} ), \quad \Sigma^+ = \frac{1}{\sqrt{2}} (\Sigma^2 + \im \Sigma^3) = \sqrt{2} \im l \wedge m, \quad \Sigma^- = \frac{1}{\sqrt{2}}( \Sigma^2 - \im \Sigma^3) = \sqrt{2} \im n \wedge \bar{m}
\end{align}
Note that, importantly, all these 2-forms are complex, and, unlike in the Riemannian case, $\Sigma^\pm$ are no longer related by the complex conjugation. These 2-forms satisfy the following algebraic conditions
\begin{gather}\label{product-algebra}
	\Sigma^\pm \cdot \Sigma^1= \mp\im \Sigma^\pm, \quad \sd^1 \cdot \sd^\pm = \pm \im \sd^\pm, \quad \Sigma^1 \cdot \Sigma^1 = -\mathbb{I}, \\
	\Sigma^\pm \cdot \Sigma^\pm = 0, \quad \sd^\pm \cdot \sd^\mp = \mp \im \sd^1 -\mathbb{I}
\end{gather}
For completeness, their $\Sigma: T^*M\to T^*M, V_\mu \to \Sigma_\mu{}^\nu V_\nu$ action on the null 1-forms is as follows
\begin{align}
	\begin{array}{c|c|c|c}
		& \sd^1 & \sd^+ & \sd^- \\
		\hline 
		l & -\im l & 0 & \sqrt{2} \im \bar{m} \\
		n & +\im n & \sqrt{2} \im m & 0 \\
		m & -\im m & 0 & \sqrt{2} \im n \\
		\bar{m} & +\im \bar{m} & \sqrt{2} \im l & 0
	\end{array}
\end{align}

Instead of defining the 2-forms $\Sigma^1, \Sigma^\pm$ by their expressions (\ref{sigmas-null-tetrad}) in terms of null tetrads, we can define
\begin{align}
	\Sigma^1 = L^i \Sigma^i, \quad \Sigma^+ = S^i \Sigma^i, \quad \Sigma^- = \bar{S}^i \Sigma^i,
\end{align}
where $\Sigma^i$ are 2-forms satisfying (\ref{eq:reality-cond}), and we have introduced a null (complex) basis in $\R^3$ satisfying
\begin{align}
	L^i L^i = 1,\quad S^i S^i = 0,\quad S^i \bar{S}^i = 1,\quad L^i S^i = 0.
\end{align}
Here $L^i\in \R^3$ is a real vector, and $\bar{S}^i$ is the complex conjugate of $S^i$. A possible choice for these vectors is
\begin{align}\label{L-S}
	L^i = (1,0,0), \quad S^i = \left(0,\frac{1}{\sqrt{2}},\frac{\im}{\sqrt{2}}\right).
\end{align}
These vectors satisfy the following identities,
\begin{align} \label{eq:null-internal-vector-identities}
	\delta^{ij} = L^i L^j + S^i \bar{S}^j + \bar{S}^i S^j, \\ \label{vector-product}
	S^i \epsilon^{ijk} = \im L^j S^k - \im L^k S^j \\ \nonumber
	\bar{S}^i \epsilon^{ijk} = \im \bar{S}^j L^k - \im \bar{S}^k L^j \\ \nonumber
	L^i \epsilon^{ijk} = \im S^j \bar{S}^k - \im S^k \bar{S}^j.
\end{align}
The product algebra (\ref{product-algebra}) then follows directly from \cref{eq:null-internal-vector-identities} and the quaternion algebra of $\Sigma^i$. The description involving $L^i,S^i$ also keeps the $SO(3,\mathbb{C})$ symmetry manifest.

\section{Some additional properties of the K\"ahler metric}

The purpose of the next subsections is to collect some properties of the K\"ahler metric, which we will need for calculations in the main text. 

\subsection{Coordinate Separability}

For the derivations of the wave operator in the main text, it will be useful to see how the metric (\ref{kerr-metric}) produces a wave operator that is separable in the $r,q$ coordinates. A computation shows that the inverse metric can be written as
\begin{align}\label{inverse-metric}
	(r^2+a^2 q^2) g_{Kerr}^{-1} = g^{-1}_1(r) + g^{-1}_2(q)
\end{align}
where
\begin{align}
 g^{-1}_1(r) = - \frac{1}{C} ((r^2+a^2) \partial_t + a \partial_\phi ){}^2 + C \partial_r^2 \\
 g^{-1}_2(q) = D \partial_q^2 + \frac{1}{D} (a(1-q^2) \partial_t + \partial_\phi)^2.
\end{align}
It is worth noting that, in view of (\ref{volume-form}), the left-hand side of (\ref{inverse-metric}) is $\tilde{g}_{Kerr}^{\mu\nu}$, the inverse densitised metric. It is very interesting to see that the inverse densitiesed Kerr metric takes a simple "separable" form. It is now easy to see that the operator $\tilde{g}_{Kerr}^{\mu\nu} \partial_\mu \partial_\nu$ is separable. Indeed, decomposing the scalar field that it can act on into Fourier harmonics with respect to $t$ and $\phi$ variables, the operator $\tilde{g}^{\mu\nu}_{Kerr} \partial_\mu \partial_\nu$ becomes a sum of two operators, one containing only functions of $r$ and derivatives with respect to $r$, and the other containing functions of $q$ and derivatives with respect to $q$. 

Let us prove (\ref{inverse-metric}). First, to condense notations we introduce
\be\label{A-B}
\rho^2 = r^2 + a^2 q^2, \quad A= dt- a(1-q^2) d\phi, \quad B= a dt- (r^2+a^2) d\phi.
\ee
Then the Kerr metric (\ref{kerr-metric}) takes the form
\be
g_{Kerr} = - \frac{C}{\rho^2} A^2 + \frac{\rho^2}{C} dr^2 + \frac{\rho^2}{D} dq^2 + \frac{D}{\rho^2} B^2. 
\ee
The dual vector fields to $A,B$ are
\be
X= (r^2+a^2) \partial_t + a\partial_\phi, \qquad Y = a(1-q^2) \partial_t +\partial_\phi.
\ee
Indeed, we have
\be
A(X) = \rho^2, \quad A(Y)=0, \quad B(X)=0, \quad B(Y)=-\rho^2.
\ee
This means that the inverse Kerr metric is given by
\be
g^{-1}_{Kerr} = - \frac{1}{C\rho^2} X^2+ \frac{C}{\rho^2} \partial_r^2 + \frac{D}{\rho^2}\partial_q^2 + \frac{1}{D\rho^2} Y^2. 
\ee
Multiplying this by $\rho^2$ we get
\be\label{g1-g2}
\rho^2 g^{-1}_{Kerr} = g_1^{-1} + g_2^{-1}, \quad g_1^{-1} = -\frac{1}{C} X^2 + C \partial_r^2, \quad g_2^{-1} = D\partial_q^2 + \frac{1}{D} Y^2,
\ee
which is (\ref{inverse-metric}). 

\subsection{An explicit formula for the Chern connection}

We will also need a more explicit formula for the connection $a$ given by (\ref{a-f}). We first write $\Sigma^1$ in the form involving 1-forms $A,B$ given by (\ref{A-B})
\be
\Sigma^1 = \im A\wedge dr - B\wedge dq.
\ee
We now raise the index of this 2-form using the metric $g^{-1}_{Kerr}$ to obtain the operator $J$
\be
J= \im A \frac{C}{\rho^2} \partial_r +\im dr \frac{1}{C} X - B  \frac{D}{\rho^2}\partial_q - dq \frac{1}{D} Y.
\ee
This operator is conformally-invariant, and so we can continue using the same expression in the K\"ahler case. We need to apply this operator to the 1-form $d \log(\sqrt{CD}\lambda_+^2)$, which only depends on $r,q$. We get
\be
a = \im  \frac{C}{\rho^2} \left( \frac{C_r}{2C} - 2\lambda_+ \right) A -   \frac{D}{\rho^2} \left( \frac{D_q}{2D} - 2\im a\lambda_+ \right)B,
\ee
where we have used $\partial_r \lambda_+ = - 2\lambda_+^2, \partial_q \lambda_+ = - \im a \lambda_+^2$. 

In the main text we need a formula for $2\im(a,dF)$, where the contraction is carried out using the metric $g_1^{-1} + g_2^{-1}$. Using the explicit expressions (\ref{g1-g2}) for the inverse metrics $g_1^{-1}, g_2^{-1}$ we have
\be
2\im g_1^{-1}(a,dF) =   \left( \frac{C_r}{C} - 4\lambda_+ \right) X F, \qquad 2\im g_2^{-1}(a,dF) =     \left( \im\frac{D_q}{D} +4 a\lambda_+ \right) Y F.
\ee
But 
\be
X - a Y= \rho^2 \partial_t, \qquad \rho^2 = \frac{1}{\lambda_+ \lambda_-}
\ee
which finally gives
\be\label{a-dF}
2\im (g_1^{-1}+ g_2^{-1})(a,dF) =  \frac{C_r}{C}  XF +\im\frac{D_q}{D} YF - 4\lambda_-^{-1} \partial_t F.
\ee

\subsection{Scalar Laplacian of the K\"ahler metric}

We now use the form (\ref{inverse-metric}) to derive a useful expression for the scalar Laplacian on the K\"ahler metric. The usual expression is
\be\label{nabla-squared}
\nabla^\mu \nabla_\mu F = \frac{1}{\sqrt{|g|}} \partial_\mu ( \sqrt{|g|} g^{\mu\nu} \partial_\nu F) = g^{\mu\nu} \partial_\mu \partial_\nu F - \Gamma^\mu \partial_\mu F,
\ee
for an arbitrary function $F$, with
\be
\Gamma^\nu =- \frac{1}{\sqrt{|g|}} \partial_\mu ( \sqrt{|g|} g^{\mu\nu}), \qquad \Gamma_\mu = g_{\mu\nu} g^{\rho\sigma} \Gamma^\nu_{\rho\sigma} .
\ee

Let us compute $\Gamma^\mu$ for the K\"ahler metric. We have
\be
\tilde{g}^{\mu\nu} = \sqrt{|g|} g^{\mu\nu} = \lambda_+^2 \tilde{g}_{Kerr}^{\mu\nu}.
\ee
We have previously computed the inverse densitiesed Kerr in (\ref{inverse-metric}). This means we have
\be
\tilde{g}^{-1} = \lambda_+^2 ( g_1^{-1} + g_2^{-1}). 
\ee
In particular,
\be
\tilde{g}^{rr} = \lambda_+^2 C, \qquad \tilde{g}^{qq} = \lambda_+^2 D.
\ee

We now compute $\Gamma^\mu$ componentwise. We have
\be
\Gamma^r = - \frac{1}{\sqrt{|g|}} \partial_r ( \lambda_+^2 C).
\ee
Using
\be
g_{rr} = \lambda_+^2 \frac{\rho^2}{C} , \qquad \sqrt{|g|} = \lambda_+^4 \rho^2,
\ee
we get
\be
\Gamma_r = g_{rr} \Gamma^r = - \frac{1}{\lambda_+^2 C} \partial_r (\lambda_+^2 C) = - 2\partial_r \log(\lambda_+ \sqrt{C}).
\ee
We similarly get
\be
\Gamma_q = g_{qq} \Gamma^q =  - \frac{1}{\lambda_+^2 D} \partial_q (\lambda_+^2 D) = - 2\partial_q \log(\lambda_+ \sqrt{D}).
\ee
This means that 
\be\label{gamma}
\Gamma = \Gamma_r dr + \Gamma_q dq = - 2 d\log(\lambda_+ \sqrt{CD}) = - 2 d\log(f).
\ee
In the last expression we wrote the result in terms of the function $f$ introduced in (\ref{f}). 

\subsection{Useful identities}

We first compute $\nabla^\mu \nabla_\mu \lambda_+$ on the K\"ahler metric. We start by computing $g^{\mu\nu} \partial_\mu \partial_\nu \lambda_+$. Because $\lambda_+$ depends only on $r,q$ we have
\be
g^{\mu\nu} \partial_\mu \partial_\nu \lambda_+ = g^{rr} \partial_r^2 \lambda_+ + g^{qq} \partial_q^2 \lambda_+.
\ee
We have
\be
\partial_r \lambda_+ = - \lambda_+^2, \qquad \partial_q \lambda_+ = - \im a \lambda_+^2,
\ee
and so
\be
\partial^2_r \lambda_+ = 2\lambda_+^3, \qquad \partial^2_q \lambda_+ = - 2a^2 \lambda_+^3.
\ee
We also have
\be
g^{rr} = \lambda_+^{-2} \frac{C}{\rho^2}, \qquad g^{qq} = \lambda_+^{-2} \frac{D}{\rho^2},
\ee
which gives
\be
g^{\mu\nu} \partial_\mu \partial_\nu \lambda_+ = \frac{2\lambda_+}{\rho^2}(C- a^2 D).
\ee
We now compare this with
\be
(d\log\lambda_+, d\log\lambda_+)= g^{\mu\nu} \partial_\mu \log\lambda_+ \partial_\nu \log\lambda_+ = g^{rr} \lambda_+^2 + g^{qq} (-a^2 \lambda_+^2) = \frac{1}{\rho^2}( C-a^2 D).
\ee
This means we have
\be\label{lambda-plus-ident}
\lambda_+^{-1} g^{\mu\nu} \partial_\mu \partial_\nu \lambda_+ = 2(d\log\lambda_+, d\log\lambda_+).
\ee

Finally, we can use it to compute $\nabla^\mu \nabla_\mu \lambda_+$. We have
\be\label{delta-lambda-plus}
\lambda_+^{-1} \nabla^\mu \nabla_\mu \lambda_+ = \lambda_+^{-1} g^{\mu\nu} \partial_\mu \partial_\nu \lambda_+ - (\Gamma, d\log\lambda_+) =
2(d\log\lambda_+, d\log\lambda_+) + 2 ( d\log(f), d\log\lambda_+).
\ee

We now compute
\be
(CD)^{-1/2} g^{\mu\nu} \partial_\mu \partial_\nu (CD)^{1/2} = C^{-1/2} g^{rr} \partial_r^2 C^{1/2} + D^{-1/2} g^{qq} \partial_q^2 D^{1/2}.
\ee
Using
\be
\partial_r^2 C^{1/2} = C^{-1/2}\left( 1- \frac{(r-M)^2}{C}\right), \qquad \partial_q^2 D^{1/2} = -D^{-1/2}\left( 1+ \frac{q^2}{D}\right),
\ee
we have
\be
(CD)^{-1/2} g^{\mu\nu} \partial_\mu \partial_\nu (CD)^{1/2} = -\frac{1}{\lambda_+^2 \rho^2} \left(\frac{(r-M)^2}{C} + \frac{q^2}{D}\right).
\ee
We can rewrite this as
\be
- \frac{\lambda_+}{\lambda_-}(CD)^{-1/2} g^{\mu\nu} \partial_\mu \partial_\nu (CD)^{1/2} = \frac{(r-M)^2}{\Delta} +\cot^2\theta.
\ee
We can rewrite the right-hand side by introducing
 \be\label{tilde-a}
 \tilde{a} = J d \log(f/\lambda_+) = J d \log\sqrt{CD} = \im \frac{C_r}{2\rho^2} A - \frac{D_q}{2\rho^2} B.
 \ee
Its square is
 \be\label{t-a-squared}
\frac{\lambda_+}{\lambda_-} g^{\mu\nu} \tilde{a}_\mu \tilde{a}_\nu = \frac{(r-M)^2}{\Delta} + \cot^2\theta,
 \ee
 and thus
 \be
 (CD)^{-1/2} g^{\mu\nu} \partial_\mu \partial_\nu (CD)^{1/2} = - (\tilde{a},\tilde{a}). 
 \ee

Another useful way of rewriting this is
\be
\left( \frac{f}{\lambda_+}\right)^{-1} g^{\mu\nu} \partial_\mu \partial_\nu \left( \frac{f}{\lambda_+}\right) = - ( d\log(f/\lambda_+),d\log(f/\lambda_+)),
\ee
where we have used the fact that $J$ is an isometry. But the left-hand side is
\begin{align}\nonumber
\left( \frac{f}{\lambda_+}\right)^{-1} g^{\mu\nu} \partial_\mu \partial_\nu \left( \frac{f}{\lambda_+}\right) &= \frac{1}{f} g^{\mu\nu}\partial_\mu\partial_\nu f - \frac{1}{\lambda_+ } g^{\mu\nu}\partial_\mu\partial_\nu \lambda_+ - 2(d\log(f),d\log(\lambda_+)) +2 ( d\log(\lambda_+),d\log(\lambda_+))  
\\ \nonumber
 &=\frac{1}{f} g^{\mu\nu}\partial_\mu\partial_\nu f - 2(d\log(f),d\log(\lambda_+)),
\end{align}
where we have used (\ref{lambda-plus-ident}). Combining with the right-hand side we get
\begin{align}
\frac{1}{f} g^{\mu\nu}\partial_\mu\partial_\nu f =- (d\log(f),d\log(f))+4 (d\log(f),d\log(\lambda_+)) -( d\log(\lambda_+),d\log(\lambda_+)).
\end{align}
Using
\be
f^{-1} g^{\mu\nu}\partial_\mu\partial_\nu f + 2 (d\log(f), d\log(f)) = f^{-1}g^{\mu\nu}\nabla_\mu\nabla_\nu f,
\ee
as well as (\ref{lambda-plus-ident}), we get an identity that we will write in the following most useful for our purposes form
\begin{align}\label{delta-f-ident}
f^{-1} g^{\mu\nu}\nabla_\mu\nabla_\nu f -& 2 (d\log(f),d\log(f)) \\ \nonumber
&= - ( d\log(f),d\log(f)) + 4(d\log(f),d\log(\lambda_+))-  ( d\log(\lambda_+),d\log(\lambda_+)).
\end{align}

From (\ref{s-delta-f}) we also have
	\begin{align}	\label{f-delta-f}
	s & = -2 \nabla^\mu \nabla_\mu \log(f) - 2 \nabla^\mu \nabla_\mu \log(\lambda_+) 
	\\ \nonumber
	& = -2 f^{-1} \nabla^\mu \nabla_\mu f - 2 \lambda_+^{-1} \nabla^\mu \nabla_\mu \lambda_+ + 2 (d\log(f),d \log(f))+ 2 (d\log(\lambda_+),d \log(\lambda_+)).
\end{align}
Substituting (\ref{delta-lambda-plus}) and (\ref{delta-f-ident}) we get an alternative useful formula for the scalar curvature. 
\be\label{s}
s = -12(d\log(f),d\log(\lambda_+)).
\ee

\section*{Acknowledgements} S.R.G. is supported by a UKRI Future Leaders Fellowship (grant number MR/Y018060/1).


\begin{thebibliography}{99}

\bibitem{Teukolsky:1973ha}
S.~A.~Teukolsky,
``Perturbations of a rotating black hole. 1. Fundamental equations for gravitational electromagnetic and neutrino field perturbations,''
Astrophys. J. \textbf{185}, 635-647 (1973)
\doi{10.1086/152444}

\bibitem{Ryan:1974nt}
M.~P.~Ryan,
``Teukolsky equation and Penrose wave equation,''
Phys. Rev. D \textbf{10}, 1736-1740 (1974)
\doi{10.1103/PhysRevD.10.1736}

\bibitem{Bini:2002jx}
D.~Bini, C.~Cherubini, R.~T.~Jantzen and R.~J.~Ruffini,
``Teukolsky master equation: De Rham wave equation for the gravitational and electromagnetic fields in vacuum,''
Prog. Theor. Phys. \textbf{107}, 967-992 (2002)
\doi{10.1143/PTP.107.967}
[arXiv:gr-qc/0203069 [gr-qc]].

\bibitem{Fackerell:1972hg}
E.~D.~Fackerell and J.~R.~Ipser,
``Weak electromagnetic fields around a rotating black hole,''
Phys. Rev. D \textbf{5}, 2455-2458 (1972)
\doi{10.1103/PhysRevD.5.2455}

\bibitem{Aksteiner:2010rh}
S.~Aksteiner and L.~Andersson,
``Linearized gravity and gauge conditions,''
Class. Quant. Grav. \textbf{28}, 065001 (2011)
\doi{10.1088/0264-9381/28/6/065001}
[arXiv:1009.5647 [gr-qc]].


\bibitem{Araneda:2016iwr}
B.~Araneda,
``Symmetry operators and decoupled equations for linear fields on black hole spacetimes,''
Class. Quant. Grav. \textbf{34}, no.3, 035002 (2017)
\doi{10.1088/1361-6382/aa51ff}
[arXiv:1610.00736 [gr-qc]].


\bibitem{Araneda:2018ezs}
B.~Araneda,
``Conformal invariance, complex structures and the Teukolsky connection,''
Class. Quant. Grav. \textbf{35}, no.17, 175001 (2018)
\doi{10.1088/1361-6382/aad13b}
[arXiv:1805.11600 [gr-qc]].


\bibitem{Green} S.~Green, "Lorenz-gauge reconstruction for Teukolsky solutions with sources in electromagnetism," Talk at the 24th Capra meeting on Radiation Reaction in General Relativity, \url{https://pirsa.org/21060044}

\bibitem{Flaherty} E.~J.~Flaherty Jr., "An integrable structure for type D spacetimes," Phys. Lett. A {\bf 46}, 391–392 (1974).

\bibitem{Derdzinski} A.~Derdzinski, "Self-dual K\"ahler manifolds and Einstein manifolds of dimension four," Compositio Mathematica {\bf 49} 405-433 (1983).

\bibitem{Krasnov:2024qyk}
K.~Krasnov and A.~Shaw,
``Kerr metric from two commuting complex structures,''
Class. Quant. Grav. \textbf{42}, no.6, 065013 (2025)
\doi{10.1088/1361-6382/adb531}
[arXiv:2408.04389 [gr-qc]].

\bibitem{Brun-Weitz} C. LeBrun, "Curvature Functionals, Optimal Metrics, and the Differential Topology of 4-Manifolds," in Different Faces of Geometry, 
Donaldson, Eliashberg, and Gromov, editors, Kluwer Academic/Plenum, 2004.

\bibitem{Moroianu} A.~Moroianu, "Lectures on Kähler Geometry," \url{https://arxiv.org/abs/math/0402223}.

\bibitem{Krasnov:2010olp}
K.~Krasnov,
``Plebanski Formulation of General Relativity: A Practical Introduction,''
Gen. Rel. Grav. \textbf{43}, 1-15 (2011)
\doi{10.1007/s10714-010-1061-x}
[arXiv:0904.0423 [gr-qc]].

\bibitem{Krasnov:2020lku}
K.~Krasnov,
``Formulations of General Relativity,''
Cambridge University Press, 2020,
ISBN 978-1-108-67465-2, 978-1-108-48164-9
\doi{10.1017/9781108674652}


\bibitem{Plebanski:1976gy}
J.~F.~Plebanski and M.~Demianski,
``Rotating, charged, and uniformly accelerating mass in general relativity,''
Annals Phys. \textbf{98}, 98-127 (1976)
\doi{10.1016/0003-4916(76)90240-2}

\bibitem{Aksteiner:2022bwr}
S.~Aksteiner and B.~Araneda,
``K{\"a}hler Geometry of Black Holes and Gravitational Instantons,''
Phys. Rev. Lett. \textbf{130}, no.16, 161502 (2023)
\doi{10.1103/PhysRevLett.130.161502}
[arXiv:2207.10039 [gr-qc]].



\end{thebibliography}
\end{document}